\documentclass{aastex701}

\usepackage{siunitx}
\usepackage{enumitem} 
\usepackage{booktabs} 
\usepackage{amsmath}
\usepackage{graphicx}
\usepackage{comment}
\usepackage{array}
\usepackage{rotating}
\usepackage{placeins}
\usepackage{soul}
\usepackage{subcaption}
\usepackage{xspace}

\newcommand{\Chandra}{{\it Chandra}}
\newcommand{\wavlex}{{\it WavLEX}\xspace}
\newcommand{\wavdetect}{{\texttt{wavdetect}}\xspace}

\newcommand{\fluxcgs}{erg~cm$^{-2}$~s$^{-1}$}

\usepackage [english]{babel}
\usepackage [autostyle, english = american]{csquotes}
\MakeOuterQuote{"}

\def\simgt{\lower.5ex\hbox{$\; \buildrel > \over \sim \;$}}
\def\simlt{\lower.5ex\hbox{$\; \buildrel < \over \sim \;$}}

\ifdefined\pdfpageattr
\AddToHook{shipout/before}{\ifcsname r@tab:excerpt\endcsname\ifnum\value{page}=\getpagerefnumber{tab:excerpt}\relax\pdfpageattr{/Rotate 90}\else\pdfpageattr{}\fi\fi}
\fi

\begin{document}

\title{The Chandra Legacy Survey of the Central 4\si{\degree} $\times$ 6\si{\degree} around Sgr A*. I. 
WavLEX Source Detection Tool}

\author[0009-0002-1660-2420]{Harry Van Der Ark}
\affiliation{Columbia University, Columbia Astrophysics Laboratory, New York, NY}
\email{hcv2109@columbia.edu}

\author[0000-0002-6126-7409]{Shifra Mandel}
\affiliation{Columbia University, Columbia Astrophysics Laboratory, New York, NY}
\email{ss5018@columbia.edu}

\author[0000-0002-9709-5389]{Kaya Mori}
\affiliation{Columbia University, Columbia Astrophysics Laboratory, New York, NY}
\email{kaya@astro.columbia.edu}

\author[0000-0003-0293-3608]{Gabriele Ponti}
\affiliation{Max-Planck-Institut f\"ur extraterrestrische Physik, Garching, Germany}
\affiliation{INAF -- Osservatorio Astronomico di Brera, Merate, Italy}
\email{ponti@mpe.mpg.de}

\author[0000-0003-0494-743X]{Ian N. Evans}
\affiliation{Center for Astrophysics | Harvard \& Smithsonian, Cambridge, MA}
\email{ievans@cfa.harvard.edu}

\author[0000-0002-5069-0324]{J. Rafael Martinez-Galarza}
\affiliation{Center for Astrophysics | Harvard \& Smithsonian, Cambridge, MA}
\email{jmartine@cfa.harvard.edu}

\author[0000-0002-6089-5390]{JaeSub Hong}
\affiliation{Center for Astrophysics | Harvard \& Smithsonian, Cambridge, MA}
\email{jaesub@head.cfa.harvard.edu}

\author[0000-0003-2840-6152]{Heshou Zhang}
\affiliation{Max-Planck-Institut f\"ur extraterrestrische Physik, Garching, Germany}
\affiliation{INAF -- Osservatorio Astronomico di Brera, Merate, Italy}
\email{heshou.zhang@mpe.mpg.de}

\author[0000-0002-5082-5049]{Tong Bao}
\affiliation{INAF -- Osservatorio Astronomico di Brera, Merate, Italy}
\affiliation{Max-Planck-Institut f\"ur extraterrestrische Physik, Garching, Germany}
\email{tong.bao@inaf.it}

\author{Charles J. Hailey}
\affiliation{Columbia University, Columbia Astrophysics Laboratory, New York, NY}
\email{chuckh@astro.columbia.edu}

\author[0000-0002-1323-5314]{Jonathan Grindlay}
\affiliation{Center for Astrophysics | Harvard \& Smithsonian, Cambridge, MA}
\email{jgrindlay@cfa.harvard.edu}

\begin{abstract}
The Galactic Center is the most crowded region of the night sky, making it both scientifically rich and observationally challenging. Classifying its point sources typically requires multi-wavelength observations with high astrometric precision. X-rays are essential probes because they penetrate the extinction that limits optical observations. \Chandra\ is the only X-ray telescope with sufficient angular resolution ($\sim$0.5$^{\prime\prime}$ on-axis) to reliably crossmatch sources against longer-wavelength Galactic Center catalogs (e.g., in the near-infrared), yet deep \Chandra\ surveys have so far covered only the central $\sim 2^{\circ} \times 1^{\circ}$. We present the \Chandra\ Legacy Survey (CLS), a new \Chandra\ survey of the central $4^{\circ} \times 6^{\circ}$ of the Galaxy and its accompanying catalog of 5461 point sources across energy bands 0.5--2 keV, 0.5--8 keV, and 2--8 keV. Prior large area X-ray surveys have struggled to reliably detect faint sources in short-exposure observations against the region's strong X-ray background. To address this, we developed \wavlex (Wavelet detection and Likelihood Evaluation of X-ray sources), a fully automated, free-to-use detection tool optimized for strong-background, low-exposure observations. When applied to the same fields, \wavlex detects 51\% more sources than the \Chandra\ Source Catalog pipeline, with equal or greater reliability. This larger, more complete census will enable population studies of the region's cataclysmic variables, pulsars, X-ray binaries, and other compact objects. Because \wavlex is optimized for strong-background, low-exposure conditions in general, it is also applicable to other background- or exposure-limited X-ray surveys. The public \wavlex source code is available for use on GitHub with DOI:\href{https://doi.org/10.5281/zenodo.24669191}{10.5281/zenodo.24669191}.
\end{abstract}

\section{Introduction}\label{sec:intro}
The Galactic Bulge (GB), generally defined as $|b| \leq 5^{\circ}$ and $|l| \leq 5^{\circ}$, and the Galactic Center (GC) at its heart, host the densest stellar population and the greatest concentration of matter and astrophysical sources in the Milky Way \citep{Perez_2019, Gerhard2017}. This concentration not only provides a large sample of sources to study within the GC, but also hosts populations of astrophysical sources rarely found elsewhere in the Galaxy, most prominently X-ray binaries (XRBs) \citep{2021ApJ...921..148M}. The GC is also the most accessible sample for studying galactic centers in general. However, the GC poses many observational challenges, as the interstellar medium (ISM) and foreground sources along the line of sight obscure sources within it. As a result, many observatories are unsuited to studying the GC; optical penetration in particular is very limited, and high angular resolution is needed to distinguish crowded sources at all wavelengths.

X-ray telescopes are uniquely able to study energetic sources within the GC, as high extinction severely limits photon penetration below 2 keV. Of the currently active X-ray imaging spectrometers, the ACIS-I instrument on \Chandra\ has the highest on-axis angular resolution of $\sim$0.5$^{\prime\prime}$, making it uniquely capable of crossmatching individual point sources against the highly crowded multi-wavelength observations of the GC. Combining \Chandra\ with multi-wavelength data to study source characteristics is therefore crucial for accurately classifying point sources in this region. In particular, large populations of cataclysmic variables (CVs) are predicted to exist within the GC but remain largely unclassified \citep{2008AIPC.1010..136H}. Classifying these sources as magnetic CVs, non-magnetic CVs, or entirely different source types requires a correlation between the near-infrared (NIR) color and the X-ray hardness ratio \citep{10.1111/j.1365-2966.2009.15826.x, 1993AJ....106.1191R}. While infrared (IR) catalogs such as VVV have covered much of the Galactic Bulge, there has been no large-scale \Chandra\ X-ray survey to crossmatch with \citep{Minniti_2010}.

Previous notable \Chandra\ surveys of the GC include the Galactic Center Survey \citep[GCS;][]{2009ApJS..181..110M}, which covers only $\sim$1.6 deg$^2$; the Bulge Latitude Survey \citep[BLS;][]{2010HEAD...11.4404G}, which also covers only $\sim$1.6 deg$^2$ and is available only through the \Chandra\ Source Catalog (CSC); and the Galactic Bulge Survey \citep[GBS;][]{Jonker_2011}, which covers $\sim$8 deg$^2$ but with an average exposure of only 2 ks. Even with the sum of these surveys, the GC remains significantly under-surveyed by \Chandra, with deep coverage limited to the central $\sim 2^{\circ} \times 1^{\circ}$ and the Limiting Window/Revnivtsev field \citep{Hong_2012, Revnivtsev_2009}.

In this work, we present a point source catalog of the Chandra Legacy Survey (CLS) -- a new \Chandra\ survey of the GC -- produced in collaboration with members of the CSC team. The survey region is bounded by $\pm2^{\circ}$ in Galactic latitude and $\pm3^{\circ}$ in Galactic longitude, and it deliberately excludes the $\sim$3 deg$^2$ already covered by the higher-exposure GCS and BLS \citep{2009ApJS..181..110M, 2010HEAD...11.4404G} and the $\sim$2 deg$^2$ region around GX 3+1, a very bright source of stray light. This results in $\sim$19 deg$^2$ of coverage, with $\sim$1.65 Ms of total exposure and a mean exposure of 4.7 ks per observation pointing. These observations were taken between March 18th 2021 and October 7th 2022, with Observation IDs (ObsIDs) 23884--24233. 

Given the low average exposure and the strong astrophysical X-ray background of the GC, we developed the new \wavlex tool to maximize the CLS catalog accuracy. \wavlex is a fully automated, free-to-use source detection and likelihood evaluation tool. Similar to the CSC source-processing pipeline, which is not publicly available in the Chandra Interactive Analysis of Observations software (CIAO) \citep{2026ApJ...1005..116F}, \wavlex uses wavelet-based detection followed by Maximum Likelihood Estimation (MLE) to evaluate each candidate X-ray source. \wavlex was designed to detect sources in strong-background, low-exposure X-ray observations with improved reliability over both the CSC pipeline (\S\ref{sec:tests}) and alternative detection tools such as \wavdetect. With \wavlex, we have detected and cataloged 5461 X-ray point sources in the CLS, which will be used to further study the GC in future publications.

This paper is organized as follows. \S\ref{sec:wvlx} describes the \wavlex pipeline, including the motivation for a new detection pipeline (\S\ref{subsec:motiv}), the initial data preparation and wavelet-based detection (\S\ref{subsec:wavdet}), the preparation of data products for MLE fitting (\S\ref{subsec:dataprep}), the MLE fitting of each source to calculate a likelihood value and assign a classification (\S\ref{subsec:mle}), the correction of each source's astrometric position (\S\ref{subsec:astrometry}), and additional source properties reported by \wavlex (\S\ref{subsec:additions}). \S\ref{sec:tests} describes the testing and validation of \wavlex, including simulations of sources with the Model of AXAF Response to X-rays (MARX) to determine the flux sensitivity (\S\ref{subsec:marxflux}) and positional accuracy (\S\ref{subsec:marxastro}), observations sliced by good time interval (GTI) to translate MLE values into statistical confidence (\S\ref{subsec:slices}), and blind surveys visually comparing the source detection effectiveness of \wavlex and the CSC (\S\ref{subsec:survey}). \S\ref{sec:results} presents the results and discussion, including the CLS source catalog (\S\ref{subsec:catalog}) and a comparison against the CSC (\S\ref{subsec:csc}). Our conclusions are presented in \S\ref{sec:concl}. Miscellaneous "edge case" tests are in Appendix \ref{sec:AB}, and discussions on a 1.2--2 keV medium-band catalog are in Appendix \ref{sec:AC}. The public \wavlex source code is available for use on GitHub with DOI:\href{https://doi.org/10.5281/zenodo.24669191}{10.5281/zenodo.24669191}.

\section{\wavlex Source Detection and Evaluation pipeline}\label{sec:wvlx}

In this Section, we describe the new \wavlex pipeline for X-ray source detection and evaluation. 

\subsection{Motivation for a New Detection Tool}\label{subsec:motiv}
Prior to the development of \wavlex, \Chandra{} catalogs typically used wavelet-based detection algorithms such as \wavdetect \citep{Freeman2002} to produce source lists \citep{Brandt_2001, Liu_2010, kumawat2025comprehensiveanalysisxraysources}. \wavdetect works by correlating count distributions within an observation image to ``Mexican Hat'' wavelet functions of different length scales, as parametrized by Equation \ref{eq:mexican_hat}. 
\begin{equation}
W'_m\!\left(\frac{x}{\sigma_x}, \frac{y}{\sigma_y}\right)
= \frac{1}{2\pi\sigma_x\sigma_y}
\left[2 - \frac{x^2}{\sigma_x^2} - \frac{y^2}{\sigma_y^2}\right]
e^{-\frac{x^2}{2\sigma_x^2} - \frac{y^2}{2\sigma_y^2}}
\label{eq:mexican_hat}
\end{equation}
Here, $W'_m(x/\sigma_x,\, y/\sigma_y)$ is the Mexican Hat mother wavelet amplitude at pixel offsets $x$ and $y$ from its center, and $\sigma_x$ and $\sigma_y$ are the scale lengths in each direction. It has a positive core surrounded by a negative annulus. Sliding it across the image scores how well the local counts match this shape, and locations whose score exceeds a significance threshold are classified as sources. This is a fast and effective method for creating a source list, especially because \wavdetect is prepackaged with CIAO. Previous \Chandra{} surveys of the GC
(e.g., \cite{2009ApJS..181..110M, Jonker_2011}) 
have used \texttt{wavdetect}; however, because \wavdetect correlates all sources with the different scales of the same model, it ignores source-specific characteristics, which limits detectability, so detecting more faint sources necessarily means accepting more spurious detections. For the purposes of studying the point source population of the GC, which requires detecting many faint sources in short ($\sim$5~ks), strong-background exposures, \wavdetect's compromise is very limiting, and it is not sufficiently reliable or comprehensive for this task.
 
The primary limitations of \wavdetect are that it considers the point spread function (PSF) size but not PSF shape, and does not fit a source-specific background. An extended PSF can blend a source into the background, and a strong background can produce spurious signals that appear similar to faint sources. The CSC has considerably improved upon the reliability of \wavdetect through its source detection pipeline, which first uses \wavdetect to identify source \emph{candidates}. Each source candidate is then evaluated with an MLE algorithm, introduced in CSC 2.0 (\citet{Evans_2024_ApJS}), which fits both a background model and a source model to the source candidate to calculate a likelihood estimate \citep{Guillaume:98}. Unfortunately, the CSC pipeline software is not publicly available through CIAO, and CSC 2.2 does not include observations made public after 2023 \citep{Evans_2024_ApJS}. More significantly for this survey, the CSC pipeline -- which is applied over regions covering the entire sky -- processes more than 15,000 observations uniformly, including observations with very high exposure, observations with minimal astrophysical background, and ACIS-S observations \citep{Evans_2024_ApJS}. Thus, while it is generally a very reliable tool, the CSC pipeline may miss some faint sources or conflate real and spurious detections in strong-background, low-exposure observations, where source detection is most challenging (\S\ref{subsec:csc}).
 
\begin{figure}
    \centering
    \includegraphics[width=1\linewidth]{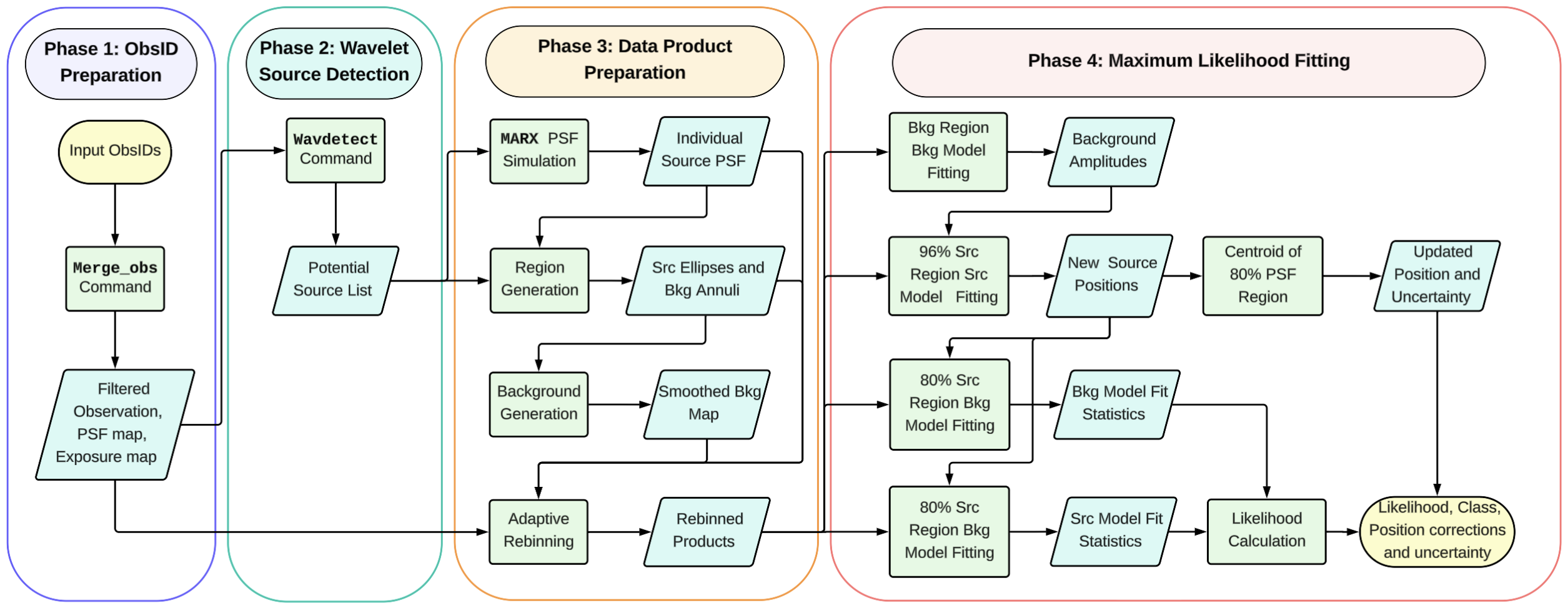}
    \caption{\wavlex flowchart showing the major processes described in \S\ref{subsec:wavdet}--\ref{subsec:mle}. Some processes, such as those in \S\ref{subsec:astrometry} and \S\ref{subsec:additions}, are omitted or simplified.}
    \label{fig:flowchart}
\end{figure}

To best analyze these strong-background, low-exposure observations, we developed \wavlex (Wavelet detection and Likelihood Evaluation of X-ray sources). \wavlex is a fully automated, public tool based on the sophisticated methods of the CSC pipeline but optimized for observations like those that make up the CLS. Figure \ref{fig:flowchart} shows a flowchart of the major phases and functions in \wavlex, with each phase discussed in \S\ref{subsec:wavdet}--\ref{subsec:mle}.

\subsection{Observation Preparation and Wavelet Source Detection}\label{subsec:wavdet}

\wavlex is fundamentally an image-analysis tool, so event files must first be converted to images before further processing. The open-source \wavlex code requires a list of ObsIDs and an energy range. For the CLS, we use three energy bands: \textit{soft} (0.5--2~keV), \textit{hard} (2--8~keV), and \textit{broad} (0.5--8~keV). The hard band is intended to utilize the high extinction within the GC, which suppresses soft photons and excludes soft foreground sources. The soft band contains foreground sources and sources with low extinction. The broad band maximizes sensitivity, recovering sources too faint to be detected in either the soft or hard band alone.
 
Phase 1 in Figure \ref{fig:flowchart} produces image files and the associated exposure maps and PSF maps for each input ObsID. It uses the CIAO commands \texttt{download\_chandra\_obsid} to download the observation packages, \texttt{chandra\_repro} to reprocess the level 2 event files, and \texttt{merge\_obs} to produce image, exposure map, and PSF map files filtered to the specified energy range. We do not use commands such as \texttt{lc\_clean} to filter out background flares, as for these low-exposure observations, shortening the good time interval (GTI) to remove flares can reduce detectability more than the lower background improves it. These files are used throughout the rest of the pipeline.

In Phase 2 of Figure~\ref{fig:flowchart}, wavelet-based source detection produces an initial list of potential sources. \wavdetect takes the image, exposure map, and PSF map files from Phase 1; the PSF map provides the PSF size as a function of off-axis angle, and the exposure map outlines the CCD edges and readout streaks, where detectability is compromised. We run \wavdetect with wavelet scales of $2^{n/2}$, where $1 \leq n \leq 14$, providing a set of scales that can capture point sources at a comprehensive range of off-axis angles, up to $\sim11'$ for ACIS-I. The significance threshold, which represents the Poisson probability of a spurious detection per pixel, is permissively set to $10^{-5}$  \citep{CXC_wavdetect_ahelp, Freeman2002}. This corresponds to $\sim10$ spurious detections per $1024\times1024$ pixel CCD. Across 350 ACIS-I observations (four CCDs each), this predicts $\sim14{,}680$ spurious detections, yet only 10,311 source candidates were detected in the broad-band total, many of them real (\S\ref{subsec:catalog}); the practical false-detection rate is therefore substantially lower than this nominal value. Previous \texttt{wavdetect}-only surveys have used significance thresholds of $10^{-6}$ or $10^{-7}$, but these thresholds exclude the detection of many faint sources \citep{Jonker_2011}. However, increasing the significance threshold above $10^{-5}$ -- or reducing the wavelet scales below $\sqrt{2}$ -- can cause individual pixels to spuriously be detected as sources; indeed, $10^{-5}$ is the significance threshold used by the CSC pipeline.
 
The principal data products of \wavdetect are a source-list FITS file with coordinates, and a region file with ellipses drawn around each potential source. Each of these source candidates is then processed and evaluated in Phases 3 and 4.
 
\subsection{Data Product Preparation}\label{subsec:dataprep}

To evaluate the sources produced in Phase 2 (\S\ref{subsec:wavdet}), we employ an MLE framework analogous to that used in the CSC.  
The MLE requires a background model (Equation \ref{bkgmdl}), a source model (Equation \ref{srcmdl}), and a region over which it is to be fit: 
\begin{equation}
    background\ model = \rm{amplitude \times background \times exposure} \label{bkgmdl}
\end{equation}
\begin{equation}
    source\ model = \left(\frac{1}{2\pi}\left(\rm 
\text{\parbox{1.75cm}{\centering source \\ amplitude}}  \times PSF \otimes e^{-\frac{1}{2}((x-x_0)^2+(y-y_0)^2)}\right) + \rm background \times
    \text{\parbox{2.0cm}{\centering background \\ amplitude}}
    \right) \times \rm exposure \label{srcmdl}
\end{equation}

Equation~\ref{bkgmdl} is simply the background times the exposure, and Equation~\ref{srcmdl} is the PSF convolved with a $\sigma=1$~pixel Gaussian, plus the background model, all times the exposure. Phase 3 in Figure \ref{fig:flowchart} outlines the production of the data products needed for the MLE fitting. The coordinates from \wavdetect are used to run \texttt{simulate\_psf}, a command that models the PSF of each point source using MARX \citep{Wise2013MARX}. MARX uses these coordinates, the aspect solution file, and a source spectrum to model the PSF. To remain impartial to the source type, the logarithmic mean of the input energy range is used as a mono-energy in place of a complete spectrum. For soft, broad, and hard CLS bands, the mono-energies are 1.1 keV, 2.7 keV, and 4.3 keV. Both the slight differences between MARX and other ray-tracing PSF models and the use of more than 1000 rays to simulate the PSF were found to have a negligible impact on low-count sources.

\begin{figure}
    \centering
    \includegraphics[width=0.8\linewidth]{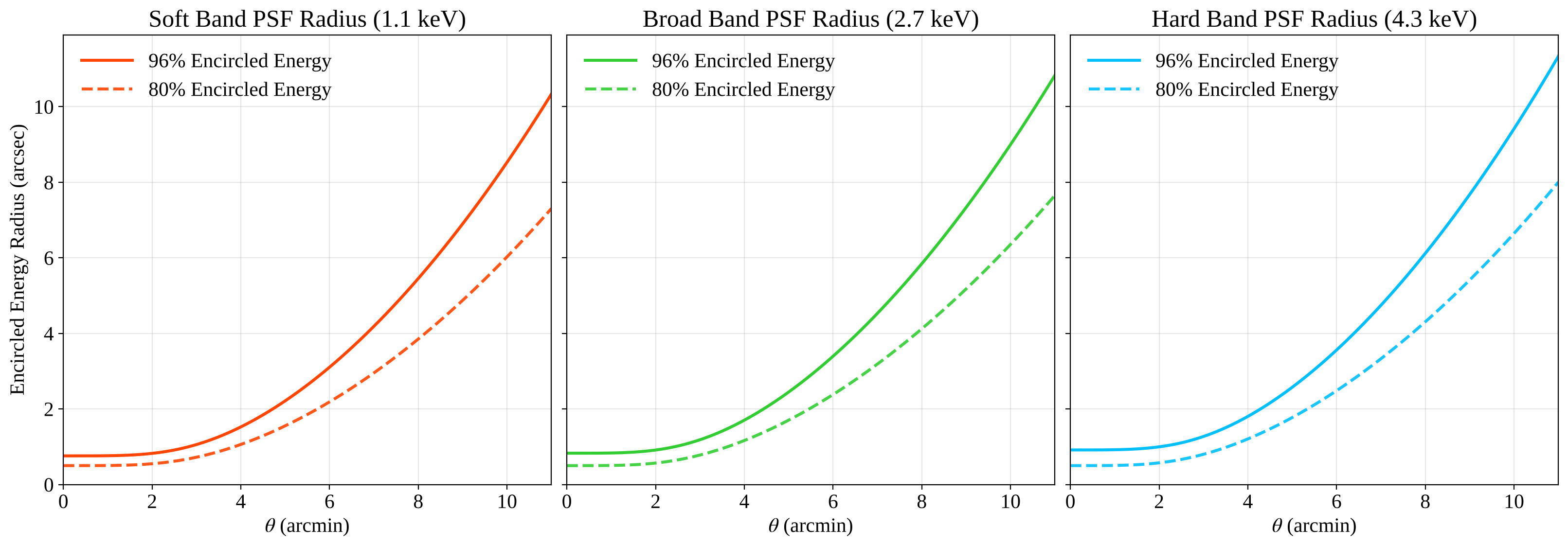}
    \caption{Radii enclosing 80\% (dashed) and 96\% (solid) of the simulated PSF flux as a function of off-axis angle $\theta$, for the soft (left), broad (center), and hard (right) bands at their mono-energies. This is a measure of the size of \wavlex regions.}
    \label{fig:psf_size}
\end{figure}

Once a PSF has been simulated for each source, the PSFs are used to produce source-fitting regions. A limitation of \wavdetect is that its source regions are based on fits to ``Mexican Hat'' functions rather than true point source PSF profiles. \wavlex instead draws elliptical regions enclosing 80\% and 96\% of the PSF flux at the mono-energy for each source, which are used in subsequent analysis. The PSF is nearly constant for $\theta\simlt2'$ but grows steeply at larger off-axis angles (Figure~\ref{fig:psf_size}): the 96\% (80\%) radius increases from $\sim0.8''$--$0.9''$ ($\sim0.5''$) on-axis to $\sim10''$--$11''$ ($\sim7''$--$8''$) at $\theta=11'$, and is larger at higher energies \citep[see also][]{CXC_POG_ch4}. A background annulus is also produced, with an inner boundary at 1.1$\times$ the 96\% PSF boundary, which leaves $\lesssim 3\%$ of source counts. The outer boundary is set to 5$\times$ the 96\% PSF boundary, with a minimum radius of 50 pixels (pixel width = $0.492''$). The region must be sufficiently large to robustly fit the background model in Phase 4. Other candidate sources that overlap the background annulus are excised using their 96\% PSF regions before the background fit (\S\ref{subsec:mle}), so that they do not bias the background amplitude. Portions of the annulus that extend beyond the detector edge are masked, and CCD edges, chip gaps, and readout streaks that intersect the regions are accounted for by the exposure map. Closely spaced sources whose regions overlap are discussed in Appendix~\ref{subsec:crowding}.

The models in Equations \ref{bkgmdl} and \ref{srcmdl} both require a background. A smoothed background map is used to capture any varying structure in an observation's background count density. To isolate the background signal, all potential sources detected by \wavdetect are removed from the image file using their 96\% PSF regions. The CIAO command \texttt{dmfilth} is then used to refill the emptied regions based on Poisson statistics and the surrounding background \citep{2026ApJ...1005..116F}. For accurate MLE fitting, the background map should be as smooth as possible while still preserving background structure; a larger smoothing scale increases the background amplitude in source regions that \texttt{dmfilth} may have underestimated. We found the best balance between underfilled regions and high background resolution for the CLS data by convolving the refilled image with a circular Gaussian of $\sigma = 50$ pixels. Exposure corrections are not applied to the background map since exposure differences were already captured in the original observation.
 
These products -- the PSF models, the PSF-defined source and background regions, and the smoothed background map -- are all that is needed to perform the MLE fitting of each source. However, we found that the likelihood of far off-axis sources was typically underestimated, as source counts are spread across more pixels, resulting in lower surface brightness. To mitigate this effect, \wavlex rebins the source image, PSF, background map, and exposure map by a factor proportional to the square root of each source's PSF area (Equation \ref{binsize}). 

\begin{equation}
    rebin\ scale = \left\lceil 0.5\times \sqrt{0.8\times(
    \rm PSF_{major\, axis} 
    \times 
    PSF_{minor\, axis})
    }\, \right\rceil \label{binsize}
\end{equation}

\begin{figure}
    \centering
    \includegraphics[width=0.5\linewidth]{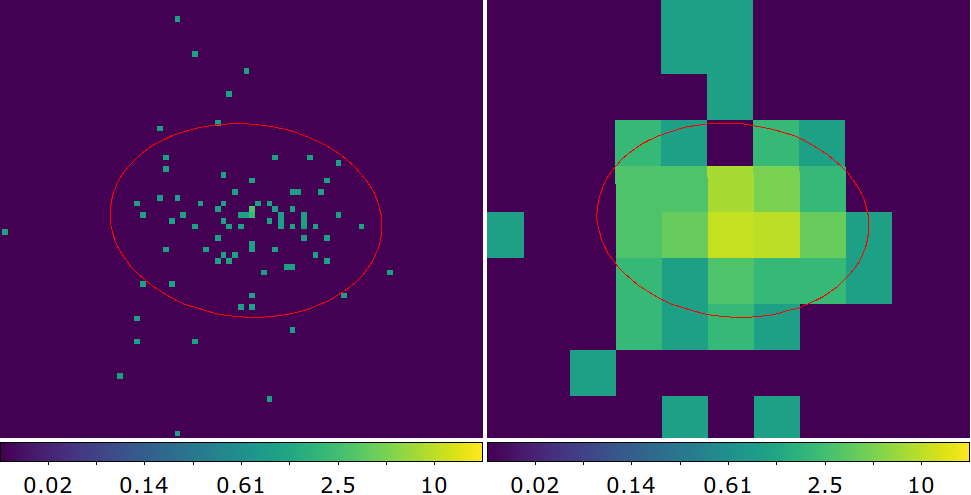}
    \caption{CXO J174517.0-321356 unbinned (left) and binned with $rebin\ scale = 8$ from Equation \ref{binsize} (right). This was the brightest source in its observation with 55 counts above the background count rate, but at $\theta=10.6'$. Unbinned $\mathcal{L}=18$, once rebinned $\mathcal{L}=173$. The red outline shows the 96\% PSF region and the color scale shows counts per pixel.}
    \label{fig:rebinned_source}
\end{figure}

We tuned the coefficient in Equation \ref{binsize} and found that a value of 0.5 best recovers the likelihood of off-axis sources, while leaving on-axis sources unaffected. In the common case where sources have a random uniform distribution across a detector, on average $\sim$50\% of sources (those at off-axis angles $\theta>6'$) are rebinned. An example of a rebinned source is shown in Figure \ref{fig:rebinned_source}.
 
\subsection{Maximum Likelihood Estimation Fitting and Classification}\label{subsec:mle}

MLE algorithms have three main steps: 
\begin{enumerate}\setlength{\itemsep}{-3pt}
    \item fitting the background model (Equation \ref{bkgmdl}) to the background,
    \item fitting the background model to the source, and 
    \item fitting the source model (Equation \ref{srcmdl}) to the source
\end{enumerate}
\citep{Guillaume:98}. The fit statistics of the source against the source model and the background model can then be used to calculate a likelihood estimate. All fitting is done using \texttt{Sherpa}, the fitting software prepackaged with CIAO \citep{2024ApJS..274...43S}.

Phase 4 in Figure \ref{fig:flowchart} begins by fitting the background model to the background annulus region produced in Phase 3 (Step 1). For this step, the background image before smoothing or refilling is used to avoid skewing the results with a bright source in the background region. The only free parameter is the dimensionless \lq background amplitude\rq\ scale factor. Once the best-fit background amplitude is determined, it is used for subsequent background modeling of the source. This step calibrates the background level in cases where the signal distribution has been altered by smoothing.

Typically, the next step of an MLE algorithm is to fit the calibrated background model to the source region (Step 2 above). Instead, \wavlex performs a source model fit using the 96\% PSF source region to improve astrometric precision in later steps (\S\ref{subsec:astrometry}). Up to this point, source coordinates were provided by \wavdetect, corresponding to the center of its wavelet model. By fitting the 96\% source region with free position parameters, both the PSF core and wings can be used to determine the most probable source center. The coordinates $x_0$ and $y_0$ of the PSF center, as well as the source amplitude, are the only free parameters of this fit. For all source fitting, we use the Nelder--Mead simplex optimization method prepackaged with \texttt{Sherpa}, which we found to be the most efficient at consistently finding the global minimum of the fit statistic given appropriate parameter limits \citep{10.1093/comjnl/7.4.308, 2024ApJS..274...43S}.

After the source position has been corrected, \wavlex performs another background model (Step 2) and source model (Step 3) fit to the source, using the re-centered 80\% PSF region with fixed $x_0$ and $y_0$. This region excludes the PSF wings, but since many of the faint sources that \wavlex is designed to evaluate have very few counts beyond the PSF core, an 80\% PSF region was empirically found to be the most reliable. Maximum likelihood calculations are based on Cash statistics \citep{1979ApJ...228..939C}, which are therefore used when fitting the source regions with \texttt{Sherpa}. The background-model and source-model Cash fit statistics are then used to compute the final likelihood with Equation \ref{likelihood}:
 
\begin{equation}
    \mathcal{L} \equiv -\log \left(\frac{\int^{\infty}_{(C_{\rm null}-C_{\rm best})/2} x^{(\nu/2) - 1}e^{-x}dx}{\int^{\infty}_0 x^{(\nu/2) - 1}e^{-x}dx}\right)\label{likelihood}
\end{equation}
where $C_{\rm best}$ is the source-model best-fit statistic, $C_{\rm null}$ is the background-model fit statistic, and $\nu$ represents the fit degrees of freedom.   Sources with likelihood $\mathcal{L} \geq 10$ are classified as ``True'', sources with $2.5 \leq \mathcal{L} < 10$ are classified as ``Marginal'', and sources with $\mathcal{L} < 2.5$ are classified as ``Unlikely''. The determination of these thresholds and the meaning of each classification are discussed in \S\ref{subsec:slices}. 

Finally, the number of counts $N$ within the 80\% PSF region is measured using the CIAO command \texttt{dmstat}; if $N<3$ or $N\leq B$ (where $B$ is the expected number of background counts), the source is classified as ``Too Faint.''  $B$ is calculated using Equation \ref{b_est}.
\begin{equation}
    B \equiv\lambda_B \times n_{pix} \times \epsilon_{norm} \label{b_est}
\end{equation}
where $\lambda_B$ is the mean number of background counts per pixel of the observation, $n_{\rm pix}$ is the number of pixels in the 80\% PSF region, and $\epsilon_{\rm norm}$ is the mean effective exposure of the source region divided by the mean effective exposure over the entire observation.  
We found that this minimum count threshold reduced spurious detections for faint on-axis source candidates and for candidates in regions of high background ("false positives").

\subsection{Source Position Alignment}\label{subsec:astrometry}
The CLS field contains an average of $\sim$800,000 NIR VVV sources per deg$^2$ \citep{Minniti_2010}, so it is important to maximize astrometric accuracy and precision for multi-wavelength crossmatching. As discussed in \S\ref{sec:intro}, \Chandra{} is best suited to this task, with the highest angular resolution of any X-ray telescope. On axis the PSF 80\% enclosed counts radius is $\sim 0.5''$, but towards the edge of the ACIS-I chip at $\sim 10'$ the 80\% enclosed counts radius significantly deteriorates to $\sim 7$--$8^{\prime\prime}$, depending on photon energy (Figure \ref{fig:psf_size}). The off-axis \Chandra{} PSF is also somewhat asymmetric, so the PSF core is offset from the centroid position. Beyond individual observations, the absolute astrometric error of \Chandra{} ACIS-I sources is $<1^{\prime\prime}$ for 90\% of sources after reprocessing \citep{CXC_absolute_astrometry}. Astrometric offsets between \Chandra{} observations can be corrected by aligning the brightest source for pointings within $1'$ of each other. The tiled CLS pointings are too widely separated for this, with a median nearest-neighbor separation of $10.4'$. Crossmatching \Chandra{} sources against Gaia can also be used for astrometric corrections, with a mean offset of $0.15''$ from stacked observations, but optical extinction makes this infeasible in the GC \citep{MartinezGalarza2022_CSC_astrometry}. For the CLS, the most feasible way to improve astrometric precision is therefore to better constrain the uncertainty with PSF modeling.

After running \wavdetect (\S\ref{subsec:wavdet}), each source candidate is assigned an RA and Dec with an associated $1\sigma$ uncertainty based on the centroid of the source \citep{CXC_Detect_manual_ch14}. However, this can be inaccurate because \wavdetect does not consider the PSF shape or asymmetry. Instead, \wavlex fits the 96\% PSF region, which contains more counts and structure from the PSF wings, to find the best-fit position at which to center the 80\% PSF region as described in \S\ref{subsec:mle}. We found that binning the image (Equation \ref{binsize}) decreased both the position offset and the computing time. The uncertainty of the best-fit position can be estimated using the \texttt{Sherpa} \texttt{get\_draws} function; however, the uncertainties often do not converge for faint sources. \wavlex therefore uses the CIAO command \texttt{dmstat}, which calculates the centroid position of the 80\% PSF region and provides the $2\sigma$ uncertainties reported in the final catalog. The PSF becomes increasingly asymmetric and less well modeled away from its center \citep{gaetz2005hrma}, and we found that the 80\% PSF region effectively compromises between retaining source counts and a symmetric profile. So while there is no universal enclosed energy fraction that defines the PSF core and wings, we refer to the 80\% PSF region as the PSF core. Asymmetry in the PSF becomes more problematic at large $\theta$, such as in ACIS-S observations. However, using the PSF core centroid, we found no evidence that asymmetry affects our results, which would appear as an inconsistent uncertainty at large $\theta$ (\S\ref{subsec:marxastro}, Figure~\ref{fig:smap}). 

\subsection{Additional Functions}\label{subsec:additions}

While the primary purpose of \wavlex is to provide an accurate source list from an observation, the pipeline also provides estimates of the photon flux and the line-of-sight (LoS) column density, $N_{\rm H}=N_{\rm HI}+2N_{\rm H_2}$. 

Starting with estimating photon flux, this procedure requires the total counts $N$ detected within the source region (\S\ref{subsec:mle}) and the expected background counts $B$ (Equation \ref{b_est}). Since $N$ includes an unknown number of background counts, we model the distribution of the net source counts $S$ with a Poisson probability distribution function, as in \citet{2007ApJ...657.1026W}, using Equation \ref{fluxpdf}:
\begin{equation}
    p(S|N,B) = \frac{(S+B)^Ne^{-(S+B)}}{\int_0^\infty(S+B)^Ne^{-(S+B)}dS} \label{fluxpdf}
\end{equation}

Taking the value of $S$ that maximizes $p(S \mid N, B)$ yields the expected source counts $S_{\rm mode}$. An approximate conversion to photon flux can then be made using Equation \ref{pflux}:

\begin{equation}
   F(S)
   \,{\rm [photons\ cm^{-2}\ s^{-1}]} =\frac{S_{\rm mode}\, [{\rm photons}]}{0.8 \bar{E}_{\rm eff}\, [\rm cm^2\ s]}, \label{pflux}
\end{equation}
where $\bar{E}_{\rm eff}$ is the mean effective exposure of the source region, the factor of 0.8 accounts for the 80\% PSF region (containing 80\% of the expected counts), and $F(S)$ is the photon flux. The maximum and minimum $S$ values of the $95.45\%$ credible interval from Equation \ref{fluxpdf} are also recorded and propagated to provide the photon flux uncertainty range.

\begin{figure}
    \centering
    \includegraphics[width=0.5\linewidth]{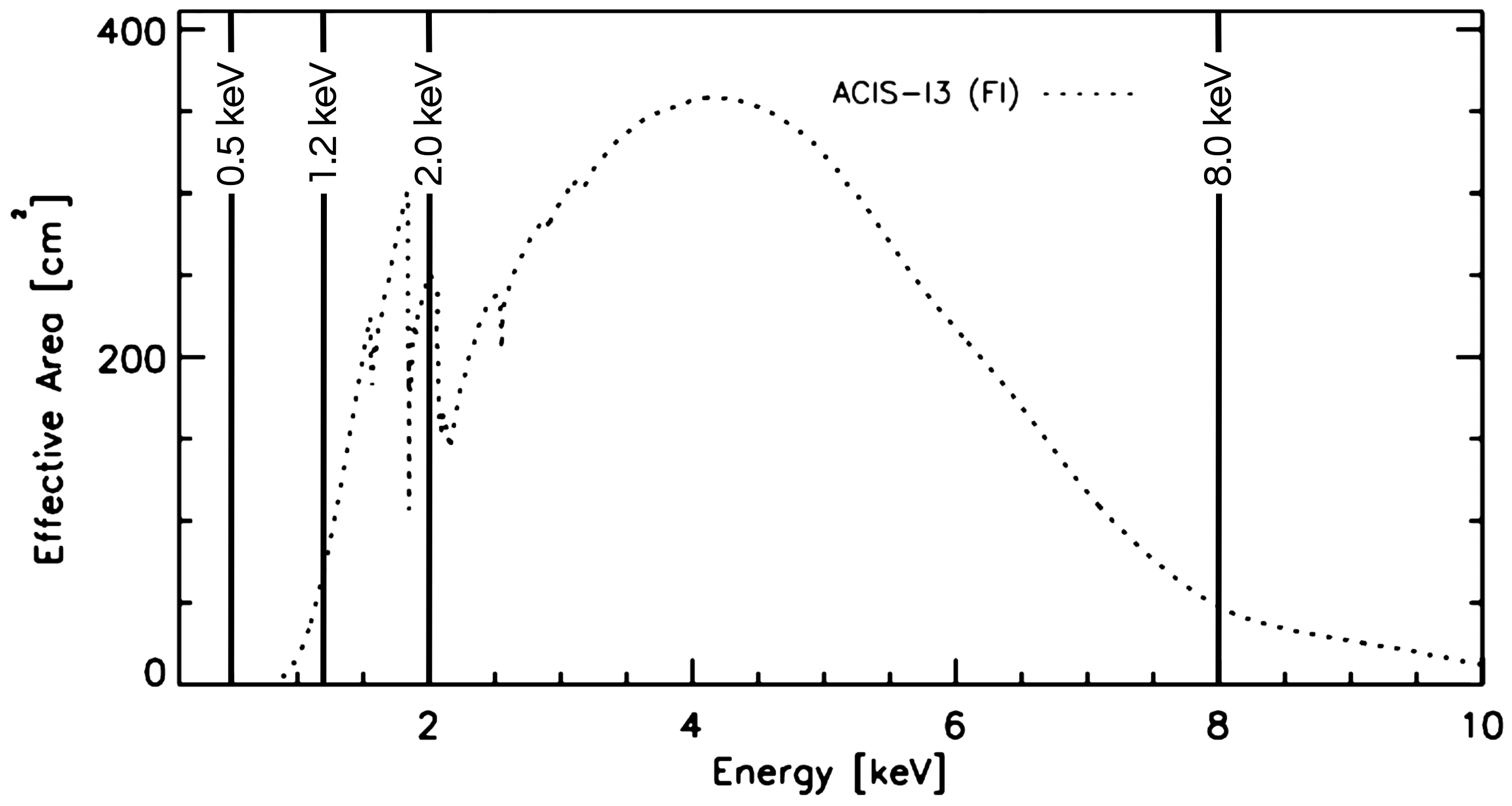}
    \caption{Effective area of a Chandra forward-illuminated ACIS-I CCD (dotted line)
as a function of photon energy. Vertical lines show the boundaries used for energy
bands in the CLS: 0.5--2.0 keV, 1.2--2.0 keV, 2.0--8.0 keV, 0.5--8.0 keV.
Adapted from the Proposers Guide from the\citet{CXC_POG}.}
    \label{fig:effarea}
\end{figure}

A limitation of not assuming a spectral model for this photon flux estimate is that the dependence of the \Chandra{} ACIS-I effective area on photon energy is not considered. Instead, for each energy band the exposure maps and PSFs are computed at a single monochromatic energy equal to the logarithmic mean of the band limits, similar to the effective energies used by the CSC \citep{csc_ebands}. As a result, sources with many high-energy photons ($>8$ keV) or low-energy photons ($<1.2$ keV) will have underestimated photon fluxes \citep{Evans_2024_ApJS,CXC_POG}. The dependence of the effective area on photon energy is shown for the ACIS-I CCDs in Figure \ref{fig:effarea}.

Beyond the photon flux, \wavlex also provides estimates of the LoS HI and H$_2$ column densities ($N_{\rm HI}$ and $N_{\rm H_2}$) for each source. While both atomic (HI) and molecular (H$_2$) hydrogen contribute to the extinction of soft X-ray sources (photon energy $<2$ keV), H$_2$ typically forms on dust grains and is thus more associated with dust scattering \citep{2017MNRAS.468.2532J, 2015MNRAS.453.1121C}.  It is especially prevalent in regions with dense molecular clouds, including the GC.

\wavlex can optionally approximate LoS $N_{\rm HI}$ for each source by comparing the source position against the HI4PI survey, a full-sky HI survey with an angular resolution of FWHM $=16.2'$ \citep{HI4PI}. Unlike for $N_{\rm HI}$, there is no current $N_{\rm H_2}$ survey covering the entire CLS; the 3-D central molecular zone (CMZ) survey by \cite{2025ApJ...984..156B} covers only $|b| < 1.4^{\circ}$. Instead, $N_{\rm H_2}$ is estimated from the CO line intensity of \citet{2001ApJ...547..792D}, which covers $|b| < 32^{\circ}$ (including the entire CLS) at an angular resolution of FWHM $=8.4'$. $N_{\rm H_2}$ is given by Equation \ref{co2h2}, where $I_{\rm CO}$ is the CO line intensity at the source position in units of [K km s$^{-1}$] and $X_{\rm CO}$ is a conversion factor that depends primarily on the radial distance of a source within the Milky Way, approximated as $5.3\times10^{19}$ cm$^{-2}$ ([K km s$^{-1}$]$^{-1}$) for the CLS \citep{Sofue_2020}:

\begin{equation}
    N_{\rm H_2} =X_{\rm CO}\times I_{\rm CO} \label{co2h2}
\end{equation}

\begin{figure}
    \centering
    \includegraphics[width=1\linewidth]{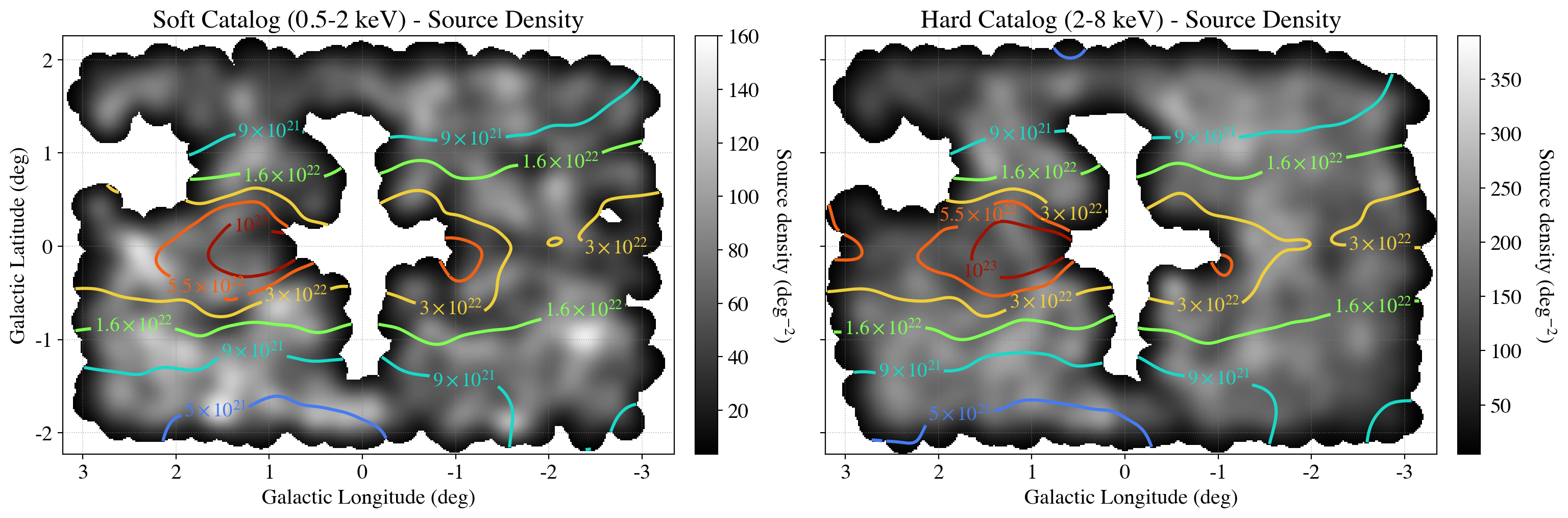}
    \caption{CLS map of source-density deg$^{-2}$ with $N_{\rm H}$ cm$^{-2}$ contours overlaid. Here $N_{\rm H} = N_{\rm HI} + 2N_{\rm H_2}$. With the soft-band catalog (0.5--2 keV; top) and the hard-band catalog (2--8 keV; bottom).}
    \label{nhcontor}
\end{figure}

Note that this estimate was derived for the CLS and may vary for other observation fields; \wavlex users who want to correct this should refer to the public GitHub. Additionally, LoS $N_{\rm H}$ is an overestimation for foreground sources. Soft sources ($<2$ keV) in particular are significantly more likely to be in the foreground except where the LoS $N_{\rm H}$ is particularly low, as shown in Figure \ref{nhcontor}.

\section{Testing and Calibrating \wavlex}\label{sec:tests}
\wavlex source detection and validation can fail in two main ways: false negatives, in which a true astrophysical source does not pass the \wavlex thresholds; and false positives, in which a spurious background fluctuation is incorrectly classified as a source. Regarding false negatives, even a perfectly designed source detection algorithm cannot reliably detect sources that are too faint to be statistically significant against the background noise without also increasing the false positive rate. It is therefore useful to quantify false negatives with a flux sensitivity function that identifies the flux above which the false negative rate is acceptably low. Regarding false positives, the likelihood $\mathcal{L}$ reported by \wavlex is already negatively correlated with the false positive rate; however, a raw likelihood value has no absolute meaning unless it is tested against detections that have been independently verified as true or false. Separately from source validation, the astrometric precision and accuracy of the reported \wavlex coordinates must also be tested.

The development of \wavlex involved extensive testing of its individual components to improve the false negative rate, the false positive rate, and the astrometric accuracy. Rather than describe each of these individual tests in detail, we present only the three principal tests in this section: MARX simulations to determine the flux sensitivity (\S\ref{subsec:marxflux}) and the astrometric precision and accuracy (\S\ref{subsec:marxastro}), and trials comparing short-exposure detections to deep observations of the same field to translate likelihood values into statistical confidence (\S\ref{subsec:slices}). Together, these tests are sufficient to evaluate the effectiveness of \wavlex within the scope for which it was designed. However, since the CSC pipeline has already processed most archival \Chandra\ observations (as of its final release, CSC 2.2; \citet{Evans_2024_ApJS}), it is also pertinent to compare the CLS and CSC source lists for the observations they share. Because neither dataset has a quantitative ground truth, we use a blind visual survey of sources for a relative comparison (\S\ref{subsec:survey}). Many other "edge cases" could also be used to test \wavlex; some of the more feasible scenarios are tested in Appendix \ref{sec:AB}.

\subsection{Flux Sensitivity of the CLS Catalog}\label{subsec:marxflux}

\begin{figure}
    \centering
    \includegraphics[width=1\linewidth]{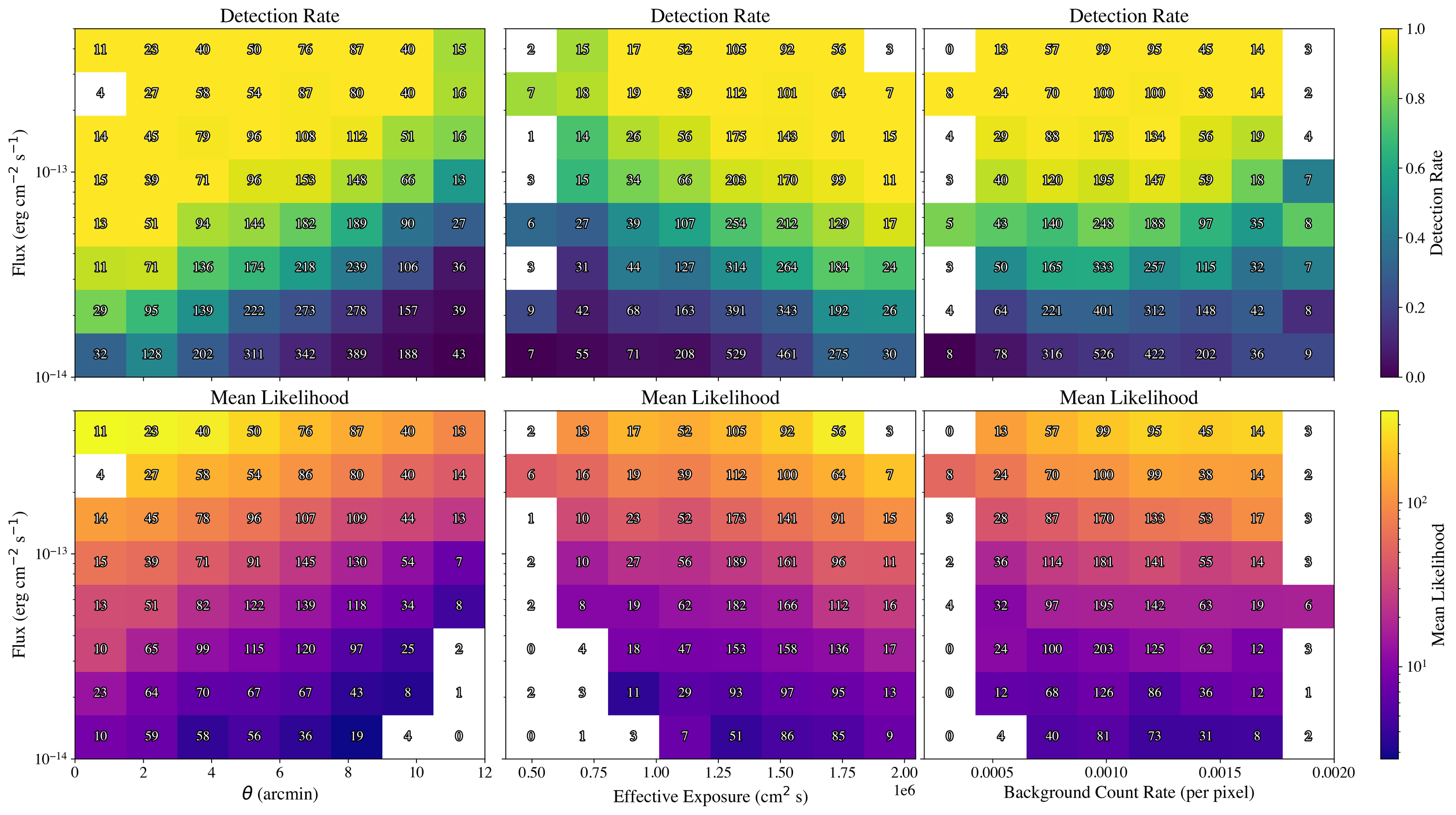} 
    \caption{Flux sensitivity heatmaps for MARX simulations similar to the CLS, as a function of off-axis angle $\theta$ (left), exposure (center), and background surface brightness (right). Rows show \wavdetect detection sensitivity (top) and \wavlex MLE likelihood scores (bottom). Each bin is marked with the number of sources within the corresponding range of flux and $\theta$/exposure/background: all simulated sources in the top row, but only those detected by \wavdetect in the bottom row, since MLE fitting is applied only to \wavdetect candidates. Color shows the detection rate (top) or the mean likelihood $\mathcal{L}$ of the detected sources (bottom). Bins containing fewer than 5 sources are excluded from each panel.}
    \label{fig:limit_heatmaps}
\end{figure}

MARX was introduced in \S\ref{subsec:dataprep} where it was used to simulate PSFs for the source model in Equation \ref{srcmdl}, although MARX can also be used to simulate many different source and instrumental parameters e.g. source spectrum and CCD edges. For testing the false negative rate, MARX is used to simulate a variety of point sources which are then injected into the \Chandra\ background of real observations within the CLS. \wavlex can be run on these simulated observations, and the recovered MARX sources can be compared to the simulation input parameters. This is the same method that was originally used to test the CSC pipeline \citep{Evans_2010}. 
For the first trials, we simulated observations with parameters similar to those of CLS observations. As in \S\ref{subsec:dataprep}, the CIAO command \texttt{simulate\_psf} is used for MARX point sources, although with notable differences. The parameter \texttt{ideal=no} is set to account for ACIS-I CCD edges and readout streaks, and instead of using a discrete energy level, a source spectrum is simulated. Most CLS sources from the GC are found within the hard band (2--8 keV), for which we model an absorbed power law with $\Gamma = 1.7, N_H = 1.2 \times 10^{22}\ cm^{-2}$. The absorbed flux of a given source is sampled from a distribution where $N \propto F_X^{-1.5}$, bounded by $[10^{-14}$, $5\times 10^{-13}$] \fluxcgs. Here  $\alpha=-1.5$ assumes uniformly distributed sources of the same luminosity and varying distance (also used by the CSC), and  $10^{-14}$ \fluxcgs\ is the minimum flux required for this model to pass the \wavlex minimum count threshold for a 5 ks exposure, as determined with PIMMS \citep{Evans_2010, Mukai1993PIMMS}. The source positions are also uniformly sampled across the detector. 32 sources were simulated per observation and injected into the sourceless background images (\S\ref{subsec:dataprep}) for 20 CLS observations; this process was repeated 10 times to provide a total of 6400 simulated sources.  \wavlex was then run on the simulated data.  The results of these simulations are shown in Figure \ref{fig:limit_heatmaps}. 

Figure~\ref{fig:limit_heatmaps} shows the flux sensitivity for our MARX simulations in terms of \wavdetect source detection rate (top) and mean likelihood $\mathcal{L}$ (bottom). Flux sensitivity depends most strongly on off-axis angle $\theta$ (left), followed by effective exposure $\bar{E}_{\rm eff}$ (center), and then background surface brightness (right). Several trends are apparent in the heatmaps:
\begin{itemize}
    \item The detection rate (top) is generally correlated with mean $\mathcal{L}$ (bottom), and both increase monotonically with flux.
    \item Both the \wavdetect and MLE threshold sensitivities correlate strongly with $\theta$ (left): the flux at which the \wavdetect detection rate reaches 50\% -- and the mean likelihood reaches $\mathcal{L}=10$ (the threshold for "True" sources) -- increases from $\sim10^{-14}$ \fluxcgs\ for on-axis sources to $\sim10^{-13}$~erg~cm$^{-2}$~s$^{-1}$ for $\theta\sim11'$. 
    \item Flux sensitivity is negatively correlated with $\bar{E}_{\rm eff}$ (center), owing to improved counting statistics at higher exposures (note, however, that CLS observations have typical exposure times of $4.7\pm0.7$~ks, hence large deviations in effective exposure occur only on the small detector areas overlapping readout streaks and CCD edges).
    \item Flux sensitivity varies little across the range of background surface brightness observed in the CLS (right).  
\end{itemize}

Therefore, when estimating the flux sensitivity in observations with comparable exposure times, off-axis angle $\theta$ is the primary parameter to consider.

Having examined one specific case ($\Gamma=1.7$, $N_{\rm{H}}=1.2\times10^{22}$~cm$^{-2}$, 2--8~keV), we now model other characteristic spectra we might expect to observe in the GC. The CLS catalog is divided into a soft-band (0.5--2~keV), a hard-band (2--8~keV), and a broad-band (0.5--8~keV). For the soft- and broad-band spectral models, characteristic power-law and blackbody parameters for GC sources were adopted from CSC~2.2 \citep{Evans_2024_ApJS}, which suit foreground sources: soft-band detections toward the GC are mainly foreground objects \citep{Muno_2003_ApJ}. The soft thermal model ($kT\approx0.2$~keV) is representative of coronally active foreground stars \citep{Mauerhan_2009_ApJ}. For the hard- and broad-band models, we used the LoS $N_{\mathrm{H}}$ for $|b|<0.8^{\circ}$ and $0.8^{\circ}<|b|<2^{\circ}$, respectively (\S\ref{subsec:additions}), since most hard X-ray sources lie near the GC. The hard model ($kT\approx8$~keV) is representative of magnetic CVs \citep{Muno_2004_ApJ_spectra}.

The characteristic soft power-law model from CSC~2.2 was highly absorbed and was therefore excluded. The spectral model parameters used in the simulations are listed in Table~\ref{tab:param}.

\setlength{\tabcolsep}{15pt}
\begin{deluxetable}{cccccccc}
\tablecaption{Spectral parameters for MARX simulations.\label{tab:param}}
\tablehead{
\colhead{Energy Band} & \colhead{$\Gamma$} & \colhead{$kT$ (keV)} & \colhead{$N_{\rm H}$ ($10^{22} \, \rm cm^{-2}$)}
}
\startdata
Soft (0.5--2 keV) &$-$ & 0.2& 0.1\\
Broad (0.5--8 keV) & 2.25 & 0.75 & 0.8\\
Hard (2--8 keV) & 1.7 & 8.0 & 3 \\
\enddata
\end{deluxetable}

Using these parameters, the soft model is simulated for soft- and broad-band data, the hard model is simulated for hard- and broad-band data, and the broad-band model is additionally simulated for data in all three energy bands. We adopt a grid of preset source parameters that samples the relevant parameter space evenly: for each model--data combination, we simulate 25 combinations of flux and off-axis angle $\theta$. Fluxes are spaced approximately logarithmically, $F_X \in \{1\times10^{-14},\,3.3\times10^{-14},\,1\times10^{-13},\,3.3\times10^{-13},\,1\times10^{-12}\}~\mathrm{erg\,cm^{-2}\,s^{-1}}$, spanning the range of feasible detection limits. Off-axis angles are $\theta \in \{0.5,\,2,\,5,\,8,\,11\}~\mathrm{arcmin}$: one near-on-axis position plus uniform $3'$ steps out to $11'$, covering the ACIS-I field of view (FOV).

\begin{figure}
    \centering
    \includegraphics[width=1\linewidth]{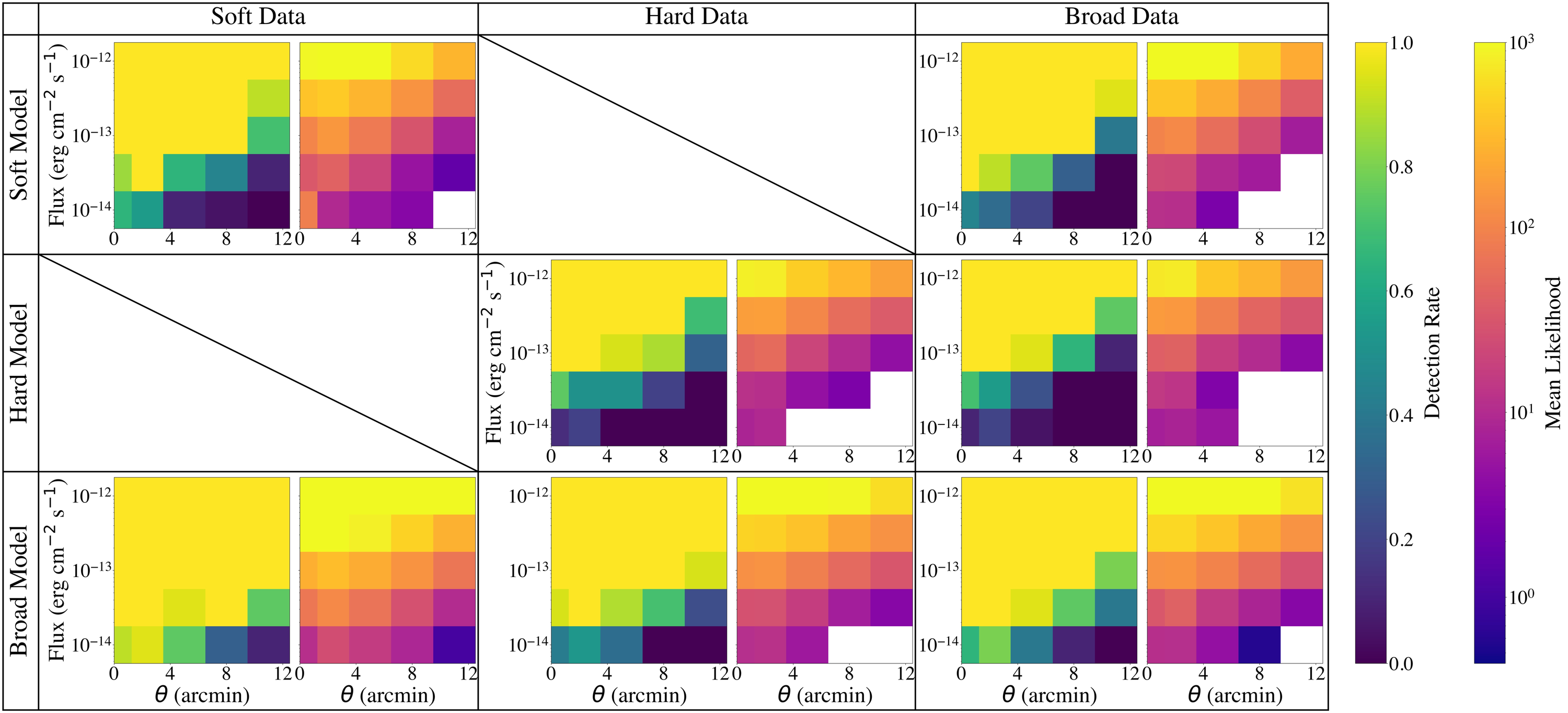}
    \caption{Flux sensitivity heatmaps for simulated absorbed blackbody models as a function of off-axis angle $\theta$. Each cell shows the \wavdetect detection rate (left) and the mean likelihood (right). Columns show the simulated energy bands, and rows show the characteristic spectral models corresponding to those energy bands. Each bin in the heatmaps represents a unique preset combination of flux and $\theta$. Certain combinations of energy band data and model are omitted because their detection in real data is unlikely. For example, only 4 sources in our crossmatch were detected in both the soft and hard bands but not in the broad band (\S\ref{subsec:catalog}).
    }
    \label{fig:bbheatmap}
\end{figure}

\begin{figure}
    \centering
    \includegraphics[width=1\linewidth]{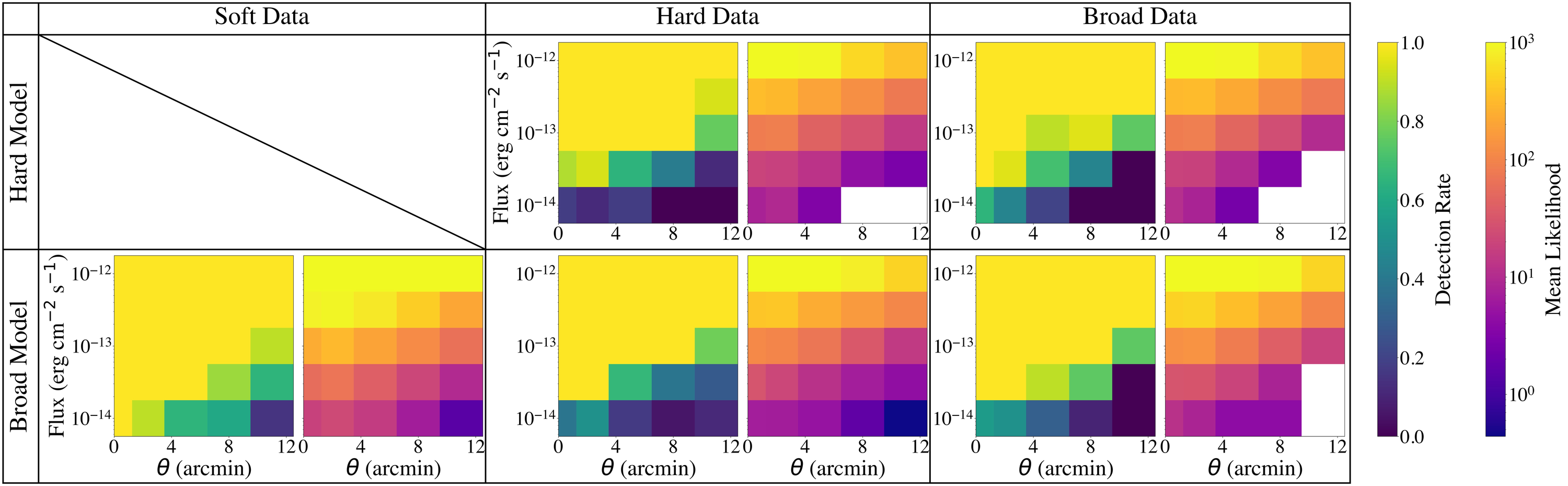}
    \caption{Same as Figure \ref{fig:bbheatmap}, but for power-law models.} 
    \label{fig:plheatmap}
\end{figure}

Note that each flux value represents the absorbed flux within the corresponding energy range. We simulated 18,000 sources in total using a sample of 20 CLS observations. The results of these simulations are shown in the heatmaps of Figures~\ref{fig:bbheatmap}--\ref{fig:plheatmap} (for the blackbody and power-law models, respectively). 


Referring to Figure \ref{fig:bbheatmap} and Figure \ref{fig:plheatmap}, the lowest flux limits are obtained for broad-band model spectra. Broad model detectability reaches the minimum simulated flux ($10^{-14}$~erg~cm$^{-2}$~s$^{-1}$) at low off-axis angles, even within the hard data where the broad model flux sensitivity was the worst; the 50\% \wavdetect detection rate and $\mathcal{L}=10$ limits stay below $(3\text{--}6)\times10^{-14}$~erg~cm$^{-2}$~s$^{-1}$ on-axis.  \wavlex appears to be least sensitive to hard-band-model sources, with detection limits degrading to about $(1.5\text{--}2)\times10^{-13}$~erg~cm$^{-2}$~s$^{-1}$ at high off-axis angles for the blackbody models (higher kT models may be more detectable in the broad-band and hard-band). In every case, sensitivity worsens by about a factor of $\sim5-10$ as we go from an off-axis angle of $0.5^{\prime}$ to $11^{\prime}$. Finally, the power-law model sources are generally detectable at fainter fluxes and larger off-axis angles than the corresponding blackbody sources, with upper limits of $(\sim5\text{--}10)\times10^{-14}$~erg~cm$^{-2}$~s$^{-1}$ (versus $(1.2\text{--}2)\times10^{-13}$~erg~cm$^{-2}$~s$^{-1}$ for the latter).

Some key insights from these simulations are that
\begin{itemize}
    \item the CLS flux sensitivity generally lies between $(1\text{--}20)\times10^{-14}$~erg~cm$^{-2}$~s$^{-1}$, with the off-axis angle $\theta$ having the greatest influence and the source spectrum, energy band filter, and effective exposure $\bar{E}_{\rm eff}$ having lesser influence. 
    \item The mean-likelihood threshold for "True" sources ($\mathcal{L}=10$; \S\ref{subsec:slices}) often occurred at flux and $\theta$ comparable to or marginally below the corresponding \wavdetect 50\% detection rate threshold. 
    \end{itemize}    
This does not mean that detection by \wavdetect is a greater benchmark of sensitivity than $\mathcal{L}=10$, because all simulated MARX sources are intended to represent real astrophysical sources, whereas not all source candidates detected by \wavdetect are real. 

What this does imply is that the combination of \wavdetect and MLE fitting in the \wavlex pipeline 
is very effective at confirming real detections. For the best case of a broad-band model in soft-band data, even very faint  ($10^{-14}$~erg~cm$^{-2}$~s$^{-1}$) sources detected by \wavdetect at $\theta\leq8.3'$ are likely to be classified as True with $\mathcal{L}\gtrsim 10$.

\subsection{Astrometric Precision and Accuracy of the CLS Catalog}\label{subsec:marxastro}
\begin{figure}[!ht]
    \centering
    \includegraphics[width=1\linewidth]{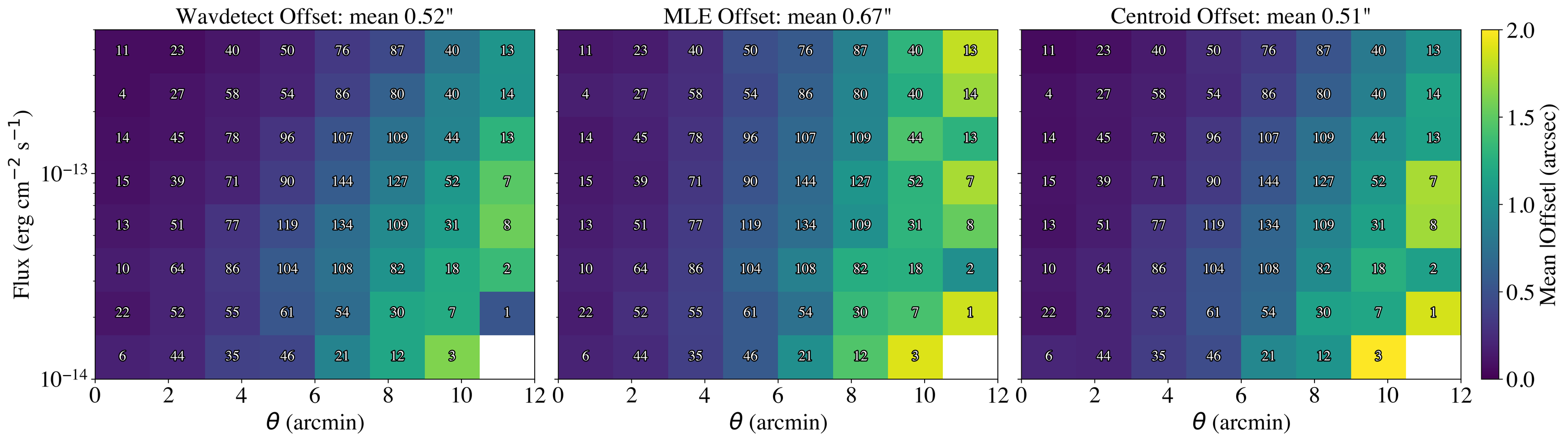}
    \caption{Heatmaps showing the mean offset of the source position from the MARX input coordinates reported by \wavdetect (left), MLE fitting (center), and the PSF core centroid (right). Offset (corresponding to bin color) is shown as a function of flux and off-axis angle $\theta$; the number of sources in each bin is also annotated. The mean offset for all sources (averaged over the range of flux and $\theta$) is reported above each heatmap.}
    \label{fig:offset}
\end{figure}

As in \S\ref{subsec:marxflux}, MARX can be used to simulate point sources to test the astrometric precision and accuracy of \wavlex. We use the same simulations employed in the first part of \S\ref{subsec:marxflux}, with 6400 sources uniformly distributed across the detector and sampled from a distribution of $N \propto F_X^{-1.5}$. With MARX, the "true" (input) coordinates RA and Dec of all simulated sources are known and can be compared with the coordinates reported by \wavlex. As discussed in \S\ref{subsec:astrometry}, \wavlex produces three sets of coordinates per source: the \wavdetect coordinates, the best-fit $x_0, y_0$ coordinates from 96\% PSF fitting (dubbed the MLE coordinates), and the centroid coordinates of the PSF core. For each of these three coordinate sets, the offset from the "true" MARX coordinates is measured, along with the \wavlex-reported $2\sigma$ position uncertainty and the percentage of MARX sources contained within the $2\sigma$ uncertainty region. The \Chandra\ PSF depends most strongly on off-axis angle, and therefore, so do the astrometric precision and accuracy (\kern-0.25em\citealp{CXC_absolute_astrometry}). Astrometric precision for the centroid models also depends on the number of source counts, so this section uses heatmaps of source flux (proportional to counts, assuming identical exposures and spectra) versus off-axis angle as an evaluation tool. The offset heatmap for each set of coordinates is shown in Figure~\ref{fig:offset}.

Figure~\ref{fig:offset} shows that for all three methods of estimating source position, the mean offset between the injected MARX and \wavlex positions is on the approximate scale of a single \Chandra\ pixel ($\sim 0.5^{\prime\prime}$), with the \wavdetect and centroid offsets being almost identical. Offsets are positively correlated with off-axis angle; the centroid and \wavdetect offsets in particular also appear weakly correlated with decreasing flux. The offset is a measure of \wavlex's astrometric accuracy, but the uncertainty of each method is also needed to understand its astrometric precision.

One could look at the reported astrometric uncertainties of \wavdetect, MLE fitting, and the PSF core centroid; however, for each method there are unknown factors that contribute to the uncertainty (e.g., the true boundaries of the PSF core) but are not incorporated into the reported errors. We model these factors by adding an unknown "noise" parameter $s$ in quadrature with the "raw" reported position variance $\sigma_{\mathrm{raw}}^2$ to obtain the "true" variance $\sigma_{\mathrm{true}}^2$ (Equation~\ref{variance}):
\begin{equation}
    \sigma_{\rm true}^2 = (1+s^2)\times\sigma_{\rm raw}^2\label{variance}
\end{equation}

\begin{figure}
    \centering
    \includegraphics[width=1\linewidth]{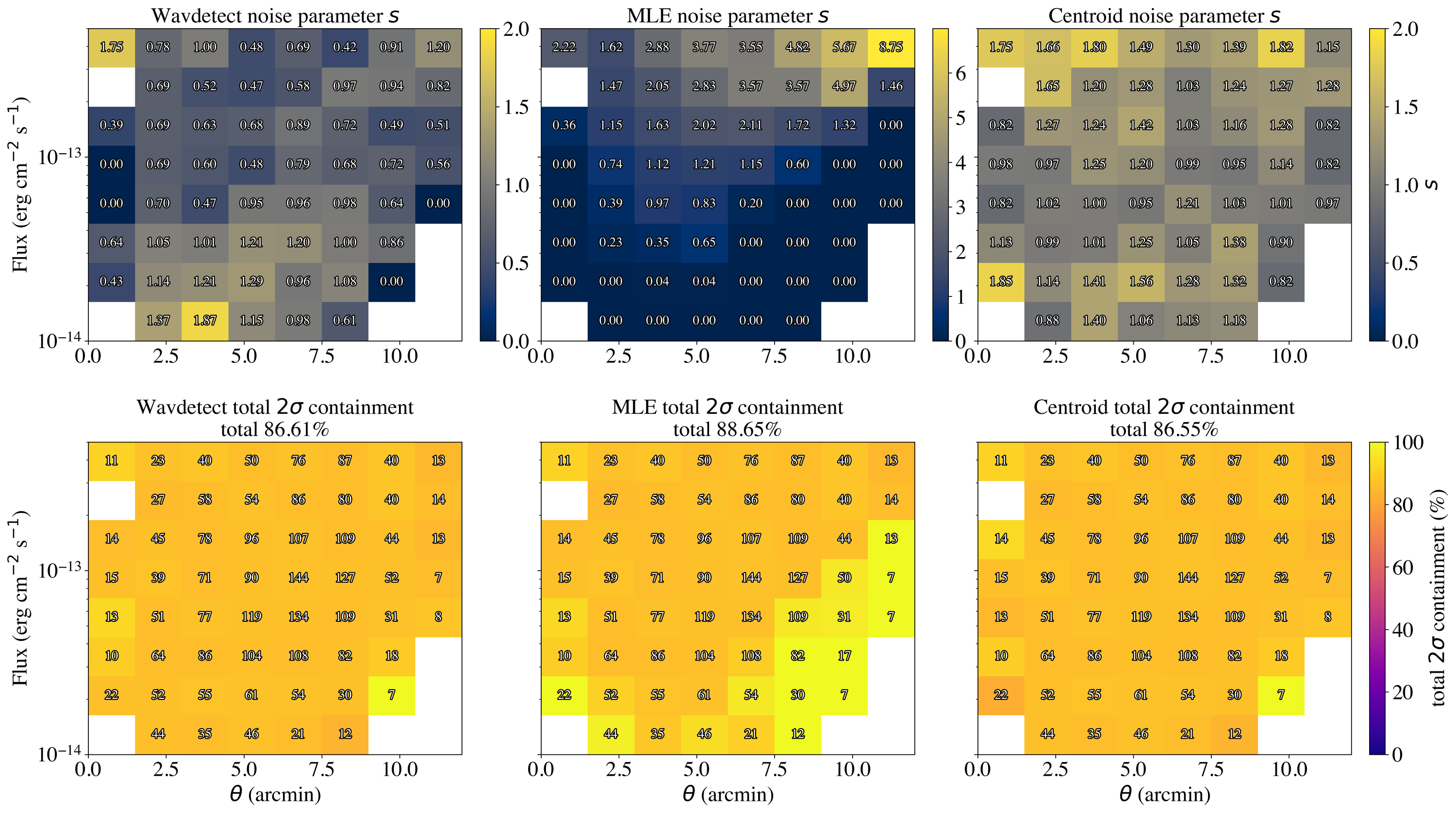}
    \caption{Heatmaps of the unknown noise parameter $s$ (top) and the percentage of MARX source positions contained within $2\sigma_{\mathrm{true}}$ (bottom). The columns show the three methods for measuring source position: \wavdetect (left), MLE fitting (center), and the PSF core centroid (right). Top: the color and annotated number of each bin give the value of $s$ required for the $2\sigma_{\rm true}$ uncertainty region to contain the input coordinates for 86.5\% of the sources. Bottom: the color of each bin gives the containment percentage using the value of $s$ from the corresponding bin above;
    the number of sources is annotated for each bin. Bins with fewer than 7 sources are removed as they cannot reach the target $\sim86.5\%$ containment.}
    \label{fig:smap}
\end{figure}

Here $s$ is a dimensionless inflation factor: $s=0$ means the reported uncertainty is already sufficient (or overestimated), while larger $s$ means $\sigma_{\rm raw}$ underestimates the true positional uncertainty by a factor of $\sqrt{1+s^2}$. It therefore measures the reliability of the reported uncertainty, not the size of the positional error itself. MARX simulations can be used to derive $\sigma_{\mathrm{true}}$ by treating $s$ as a free parameter and targeting a containment rate of $\sim 86.5\%$ for MARX source positions within a $2\sigma_{\mathrm{true}}$ error ellipse. For two fitted parameters, $\Delta\chi^2 \sim \chi^2_2$ \citep[Eq.~5, Table~1]{1976ApJ...208..177L}, whose cumulative distribution is $1 - e^{-\Delta\chi^2/2}$. The $n\sigma$ error ellipse is the contour at $\Delta\chi^2 = n^2$, so the enclosed fraction is
\begin{equation}
    \mathrm{CI}(n\sigma) = 1 - e^{-n^2/2}.
    \label{2sigmaCI}
\end{equation}
This gives $\mathrm{CI}(2\sigma) = 1 - e^{-2} \approx 0.865$, rather than the 95.4\% of the one-dimensional case. Because $\sigma_{\mathrm{raw}}$ and $\sigma_{\mathrm{true}}$ may not be linearly related across different fluxes and off-axis angles, $s$ is fit independently for each bin of flux and off-axis angle that contains at least 7 sources (Figure~\ref{fig:smap}). 

As Figure~\ref{fig:smap} (top left) shows, $s$ appears negatively correlated with flux for the \wavdetect method, indicating that $\sigma_{\mathrm{raw}}$ is increasingly underreported for fainter sources. The opposite is true for the MLE method (top center), where $s$ is strongly and positively correlated with flux. This does not mean that MLE positions are worse for bright sources; rather, it reaches an uncertainty floor based on the local minimum fit statistic, so the reported uncertainty becomes overly optimistic for bright sources. For faint sources, where the reported uncertainty is large and often poorly converged (\S\ref{subsec:astrometry}), $\sigma_{\mathrm{raw}}$ is already generous: many bins with flux below $10^{-13}$~erg~cm$^{-2}$~s$^{-1}$ reach the minimum permissible value of $s=0$, indicating that $\sigma_{\mathrm{raw}}$ may be overestimated there. For the containment heatmaps (Figure~\ref{fig:smap}, bottom), bins with $\geq7$ sources appear uniform in the percentage of sources contained within $2\sigma_{\mathrm{true}}$ (assuming the corresponding mean $s$), and the total percentage contained is within 0.1\% of the 86.5\% target, for both the \wavdetect and centroid methods. None of the methods has an average $s$ near 0 across the full range of parameters, indicating that $\sigma_{\mathrm{raw}}$ commonly underrepresents the true uncertainty. However, for the centroid method (right), $s$ does not depend on flux or $\theta$ beyond the random fluctuation expected between bins. The unknown noise term in Equation~\ref{variance}, $s\,\sigma_{\mathrm{raw}}$, is therefore proportional to the reported uncertainty, so $\sigma_{\mathrm{true}}=\sqrt{1+s^2}\,\sigma_{\mathrm{raw}}$ is linearly related to $\sigma_{\mathrm{raw}}$ across all fluxes and off-axis angles, and a single value of $s$ can be used for all bins. For the \wavdetect and MLE methods, the required $s$ varies with flux, so no such single relation holds.

\begin{figure}
    \centering
    \includegraphics[width=1\linewidth]{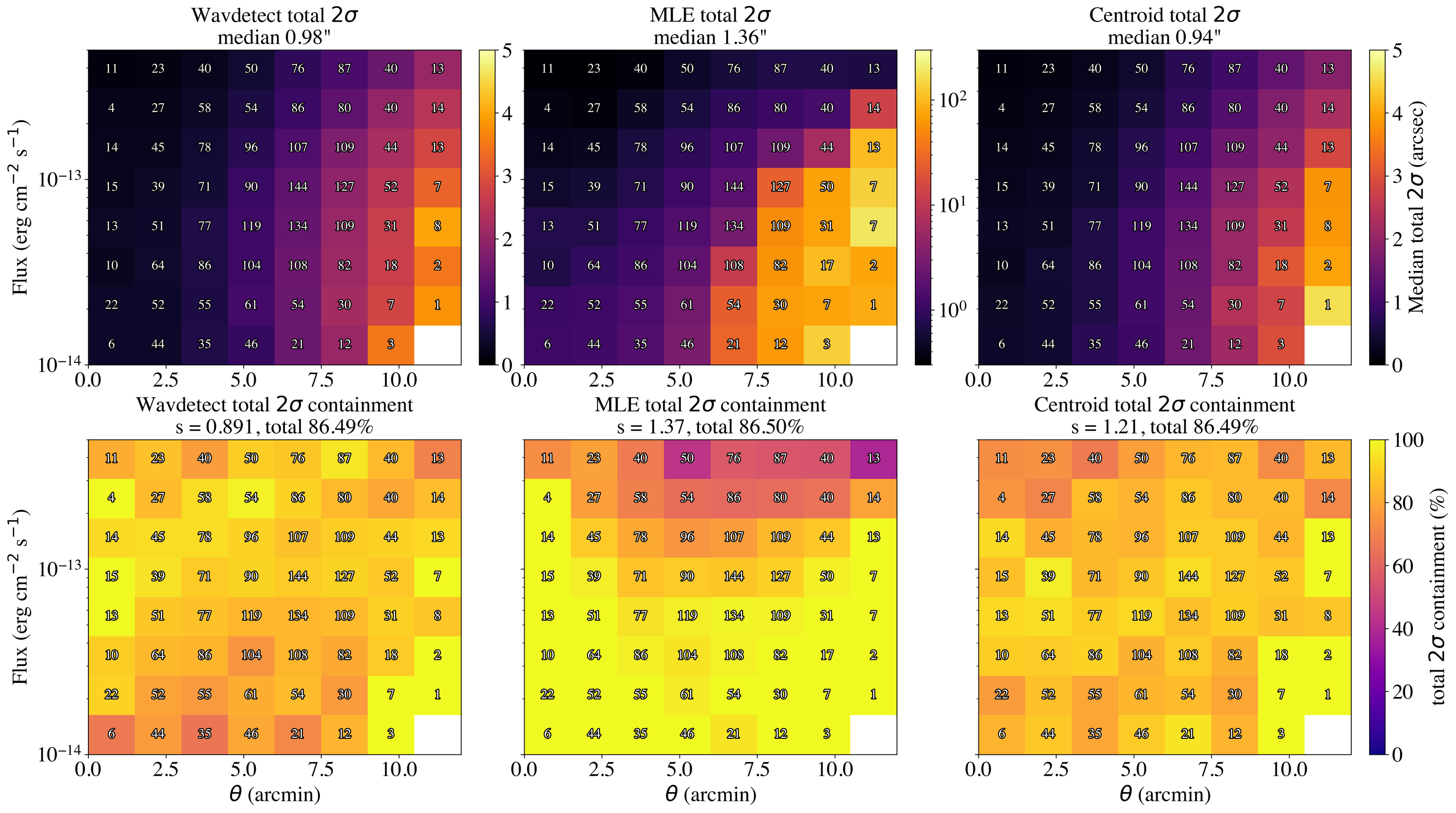}
    \caption{Heatmaps of position uncertainty $2\sigma_{\mathrm{true}}$ from Equation~\ref{variance} for constant $s$ (top row) and the percentage of MARX source positions contained within the $2\sigma_{\mathrm{true}}$ interval (bottom row). The columns show the three methods for measuring source position: \wavdetect (left), MLE fitting (center), and the PSF core centroid (right). Above each top-row heatmap is the median $2\sigma$ uncertainty across all bins, in arcseconds. Above each bottom-row heatmap is the value of $s$ used to reach containment of $\sim86.5\%$ of sources. The number of sources is annotated for each bin.}
    \label{pos_heatmap}
\end{figure}

Setting $s$ to the same value for all bins, with a target $2\sigma$ containment of 86.5\% averaged across all sources, allows $2\sigma_{\mathrm{true}}$ to be shown as a heatmap in Figure~\ref{pos_heatmap}.  

As Figure~\ref{pos_heatmap} (top) illustrates, all three methods for measuring source position have uncertainties that correlate strongly with $\theta$, as expected from the rapid growth of the PSF at $\theta>6'$. We adopted the median to show the average $2\sigma$ uncertainty; otherwise, a few sources with very large uncertainties would dominate. The centroid method has the lowest median $2\sigma$ uncertainty, $0.94^{\prime\prime}$, which is slightly below that of \wavdetect ($0.98^{\prime\prime}$) and much lower than that of the MLE method, whose uncertainty diverges to arcminute scales at large $\theta$. Although the \wavdetect and centroid uncertainties are comparable, the \wavdetect $2\sigma$ containment is correlated with flux (Figure~\ref{pos_heatmap}, bottom). This follows from the flux-dependent $s$ in Figure~\ref{fig:smap}: no single value of $s$ corrects the \wavdetect uncertainties at all fluxes, so its reported uncertainty remains less reliable across a variety of sources. Averaging the containment by flux bin results in a range of 75--91\% for \wavdetect, compared to 79--89\% for the centroid. The contribution from random fluctuations should be comparable for both, since they use the same dataset. The centroid method has marginally better accuracy and precision than \wavdetect and, more importantly, a reliable $\sigma_{\mathrm{true}}$ across a variety of sources; it is therefore the method ultimately adopted by \wavlex, with $s=1.21$ (\S\ref{subsec:astrometry}). Note that these uncertainties exclude the systematic astrometric uncertainty of the entire observation, which cannot be reliably used in the CLS fields due to the lack of overlapping observations and reliable Gaia counterparts in the Galactic Center (\S\ref{subsec:astrometry}).

\subsection{Determining Likelihood Thresholds}\label{subsec:slices}

As discussed in \S\ref{subsec:mle}, the likelihood $\mathcal{L}$ reported by \wavlex is a metric for how likely a detected source candidate is to be a real astrophysical source, but to be of practical use it must be converted into an explicit probability. \wavlex can then set likelihood thresholds that allow users to be confident in the catalog reliability. Unfortunately, the "false positive" rate (i.e., spurious source detections) cannot be tested with MARX; unlike the injection of simulated sources that can be used to test detection sensitivity, simulating background noise is far more complex. Sourceless background models do exist, such as the \Chandra\ \texttt{blanksky} files \citep{2026ApJ...1005..116F}; however, they are unsuitable for modeling the strong heterogeneous GC background. To reliably test the false positive rate, \wavlex must be run on observations with an independent set of \emph{ground-truth} sources. 

This is done using the Limiting Window (LW), a low-extinction field at $l\sim0^{\circ}$, $b\sim-1.4^{\circ}$ with $>900$~ks of stacked \Chandra\ GTI exposures \citep{Revnivtsev_2009} -- and superior detection sensitivity -- that is surrounded by CLS observations. We selected the CSC stacked source list for ACIS-I in the LW with a CSC "True" classification, broad-band flux above the \wavlex flux sensitivity, and from July-August 2008. This yielded 1158 sources, which are used as the reference "ground truth" \citep{Evans_2024_ApJS}. We then took ObsIDs 9500, 9501, 9502, 9503, 9504, 9854, 9892, and 9893 from July-August 2008, which total 801.2~ks of GTI exposures, and "sliced" them into 160 event files of 5~ks each, comparable to the average CLS exposure time. \wavlex was run on each sliced observation in the broad-band, resulting in 5906 source candidates summed from all slices that were detected by \texttt{wavdetect}, and were above the minimum count threshold.

To compare these source candidates to the ground-truth source list, we assign each candidate a search radius of $1.7^{\prime\prime}$ for $0^{\prime}\leq \theta\leq 5^{\prime}$, $3.1^{\prime\prime}$ for $5^{\prime}<\theta \leq 8^{\prime}$, and $4.5^{\prime\prime}$ for $8^{\prime}<\theta \leq 11^{\prime}$. These radii are derived from the average $2\sigma$ centroid uncertainties of $0.5^{\prime\prime}$, $1.5^{\prime\prime}$, and $2.5^{\prime\prime}$ for the respective off-axis angle ranges (\S\ref{subsec:marxastro}), which we also assume for the CSC stacked-source positions (conservative, as the deep stack should be more precise). Adding the CSC and \wavlex $2\sigma$ uncertainties in quadrature gives the $2\sigma$ uncertainty on the separation between the two positions (e.g., $\sqrt{0.5^2+0.5^2}\approx0.7^{\prime\prime}$ for $\theta\leq5^{\prime}$). We then add $1^{\prime\prime}$ which is the maximum difference between the CSC observation and stack position for a source to account for potential stack--observation offsets, which gives the search radii above. Even with these conservative assumptions, for the 1158 ground-truth sources there is only a $\sim2.4\%$ false crossmatch probability. This probability is the total area of all search circles divided by the FOV area, assuming a uniform source distribution.

\begin{figure}
    \centering
    \includegraphics[width=0.75\linewidth]{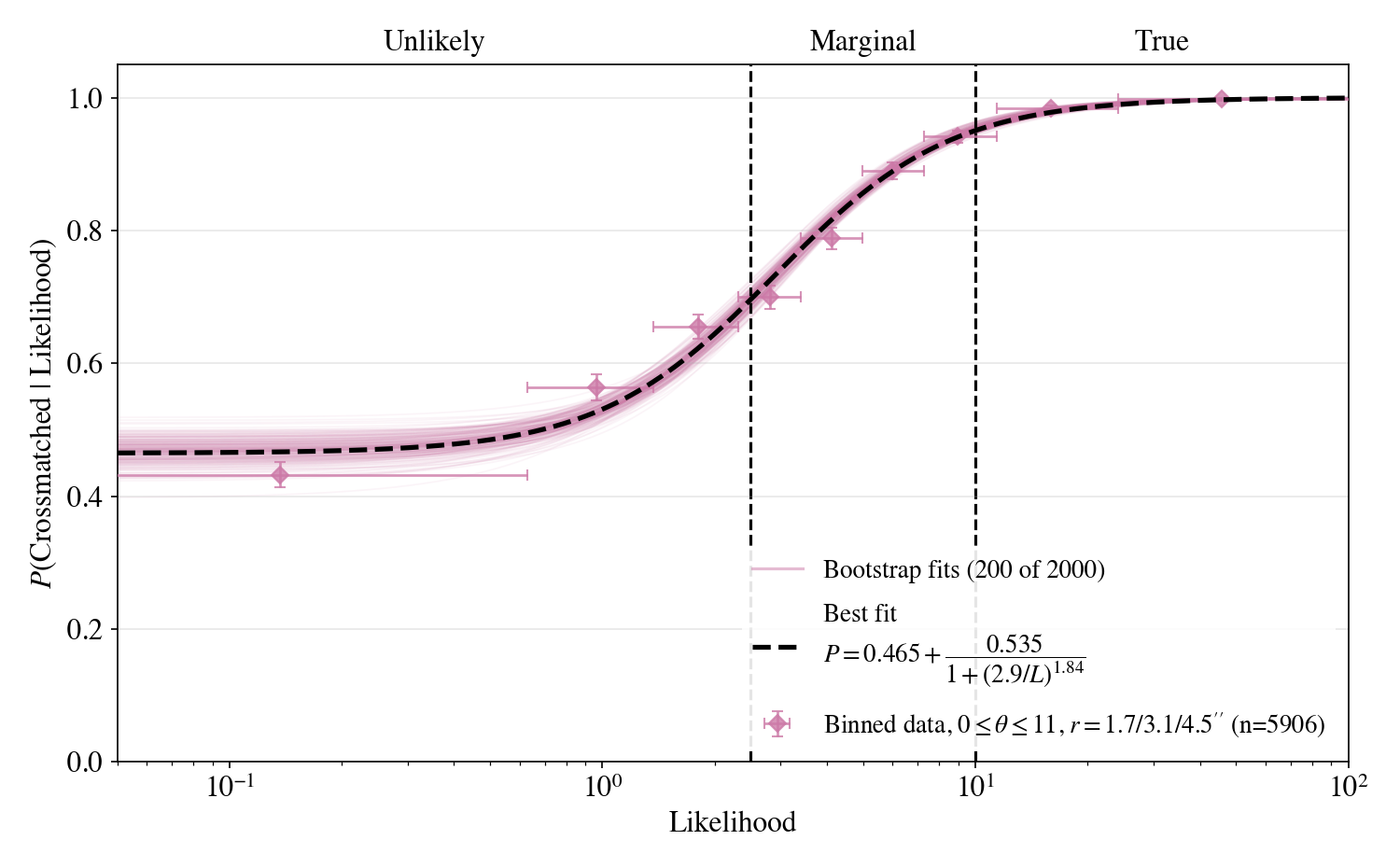}
    \caption{Crossmatch probability $P(C\mid\mathcal{L})$ as a function of \wavlex likelihood for source candidates from 5~ks slices of Limiting Window observations, crossmatched to the 800~ks stacked-observation source list. Each bin has an equal number of sources; horizontal error bars show the likelihood range of the sources in each bin, and vertical error bars show the standard deviation of $P(C\mid\mathcal{L})$. The dashed line shows the best fit of Equation~\ref{loglogistic}. Faint pink lines show the first 200 of the 2000 bootstrap realizations of Equation~\ref{loglogistic}, drawn from normal distributions around the best-fit parameters.}
    \label{fig:bootslice}
\end{figure}

The \wavlex source candidates are then binned by likelihood into 9 bins with an equal number of candidates in each. The proportion of candidates in each bin that were crossmatched with a ground-truth source gives $P(C\mid\mathcal{L})$, the crossmatch probability for a given likelihood. The results are then fit with the log-logistic function in Equation~\ref{loglogistic}:
\begin{equation}
    P(C|\mathcal{L})=\frac{1-\eta}{1+(\frac{\alpha}{\mathcal{L}})^\beta} + \eta \label{loglogistic}
\end{equation}  
where $\mathcal{L}$ is the likelihood, $\eta$ is the crossmatch probability as $\mathcal{L}\rightarrow0$, and $\alpha$ and $\beta$ control the inflection point and gradient.  
Equation \ref{loglogistic} asymptotically approaches $P(C\mid\mathcal{L})=1$ as $\mathcal{L}\rightarrow\infty$. The results of fitting Equation~\ref{loglogistic} are shown in Figure~\ref{fig:bootslice}.

Figure~\ref{fig:bootslice} shows that $P(C\mid\mathcal{L})$ increases monotonically with \wavlex likelihood, and Equation~\ref{loglogistic} fits the binned data with a $\chi^2$ of 17.34 (6 d.o.f). The model fits more poorly at low likelihood; for $\mathcal{L}\geq2$, $\chi^2=2.46$ (3 d.o.f). Additionally, the uncertainties in the binned data were used to determine uncertainties for the model parameters $\eta$, $\alpha$, and $\beta$, so that the model could be bootstrapped over 2000 iterations. The best-fit model reaches a $1\sigma$ confidence of $P(C\mid\mathcal{L})=0.683$ at a likelihood of 2.37 and a $2\sigma$ confidence of $P(C\mid\mathcal{L})=0.955$ at a likelihood of 10.6. From the bootstrap models, we determine the $1\sigma$ confidence intervals of these likelihood thresholds to be $\mathcal{L}_{1\sigma}=2.37\pm0.14$ and $\mathcal{L}_{2\sigma}=10.55^{+0.40}_{-0.93}$. These results are implemented in \wavlex by setting the threshold for \emph{Marginal} sources at $\mathcal{L}=2.5$ ($\sim1\sigma$ confidence) and the threshold for \emph{True} sources at $\mathcal{L}=10$ ($\sim2\sigma$ confidence). Equivalently, if the stacked LW list is taken as ground truth, $\sim32\%$ of candidates at the Marginal threshold ($\mathcal{L}=2.5$) and $\sim5\%$ of candidates at the True threshold ($\mathcal{L}=10$) have no stacked counterpart, so the Marginal list as a whole is expected to contain $\sim5$--$32\%$ unmatched candidates. These fractions are upper limits on the false-positive rate, since an unmatched candidate is not necessarily spurious: it may be a real source that falls below the flux cut applied to the ground-truth list or a variable source (e.g., flaring during a 5~ks slice but faint in the stacked average), although the latter would require analyzing the lightcurves to verify.

\begin{figure}[!ht]
    \centering
    \includegraphics[width=1\linewidth]{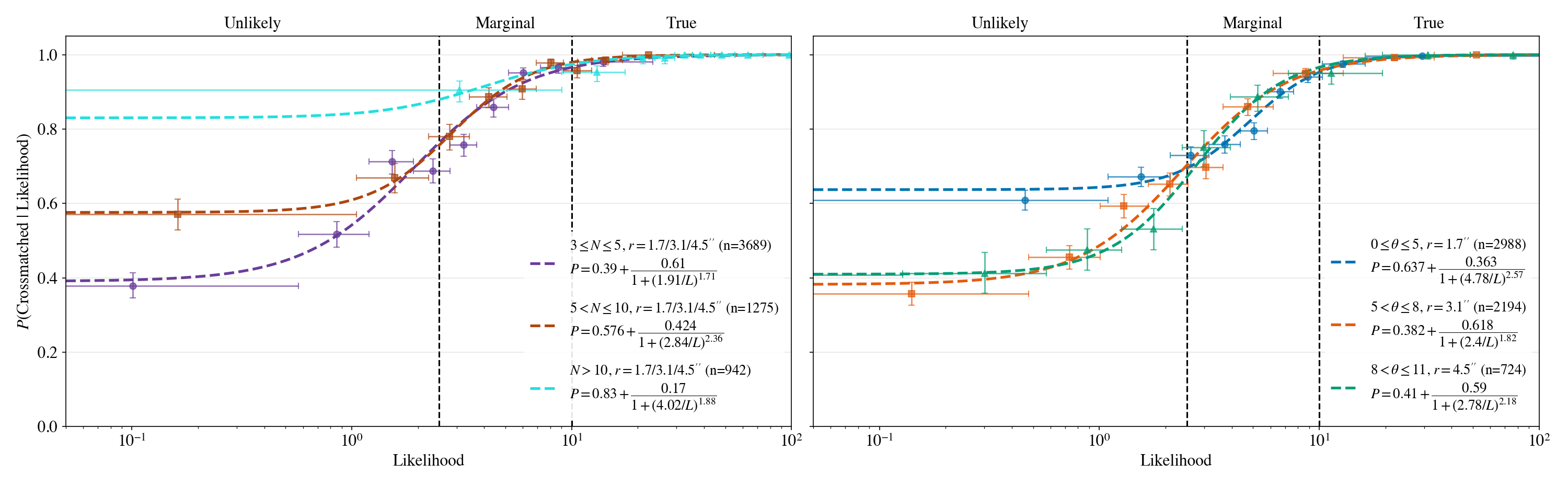}
    \caption{Crossmatch probability $P(C\mid\mathcal{L})$ as a function of \wavlex likelihood for source candidates from 5~ks slices of Limiting Window observations, crossmatched to the 800~ks stacked-observation source list. Left: source candidates grouped by number of counts. Right: source candidates grouped by off-axis angle $\theta$. Within each group, there are 9 bins with an equal number of sources (e.g. $\sim105$ sources per bin for $N>10$); horizontal error bars show the likelihood range of the sources in each bin, and vertical error bars show the standard deviation of $P(C\mid\mathcal{L})$. The number of sources in each group and the corresponding best-fit models are given in the legends.}
    \label{fig:groupslice}
\end{figure}

To determine the reliability of these likelihood thresholds, it is important to examine whether $P(C\mid\mathcal{L})$ is consistent when the source candidates are grouped by different parameters. Indeed, the CSC pipeline sets likelihood thresholds as a function of source parameters such as off-axis angle \citep{nowak2016csc2thresholds}. In Figure~\ref{fig:groupslice} (left), the sliced-observation source candidates are grouped by number of counts ($3\le N_{\mathrm{cts}}\le5$, $5<N_{\mathrm{cts}}\le10$, and $10<N_{\mathrm{cts}}$).  Figure~\ref{fig:groupslice} (right) shows the source candidates grouped by $\theta$ ($0^{\prime}\le\theta\le5^{\prime}$, $5^{\prime}<\theta\le8^{\prime}$, and $8^{\prime}<\theta\le11^{\prime}$).

From Figure~\ref{fig:groupslice}, the best-fit models of Equation~\ref{loglogistic} diverge for different source counts and off-axis angles. The horizontal error bars, which show the range of likelihood within each bin, reveal that high-count sources (cyan markers in Figure~\ref{fig:groupslice}, left) have higher typical likelihoods, whereas sources at large $\theta$ (green markers in Figure~\ref{fig:groupslice}, right) have lower typical likelihoods. This has the effect of setting $\eta\sim P(C\mid\mathcal{L})$ at higher likelihoods for brighter, on-axis sources. When comparing the $P(C\mid\mathcal{L})$ values of different groups at $\mathcal{L}=2.5$ and $\mathcal{L}=10$, where the classification thresholds are set, the groups either align closely, with overlapping error bars, or one of the groups has significantly fewer source candidates at that likelihood. The only apparent divergence is for $N_{\rm cts}>10$ near $\mathcal{L}\sim2.5$, where $P(C\mid\mathcal{L})$ is higher than for the other groups. This group has few low-likelihood candidates, so with equal-count binning its lowest-likelihood bin extends up to $\mathcal{L}\approx9$. Because $P(C\mid\mathcal{L})$ rises steeply across this range, the crossmatch fraction of this bin is dominated by its higher-likelihood members and overestimates $P(C\mid\mathcal{L})$ at its plotted position of $\mathcal{L}\sim2.5$. This implies that $P(C\mid\mathcal{L})$ is primarily a function of likelihood. Likelihood thresholds could also be tested separately for different energy bands. However, \S\ref{subsec:marxflux} showed that the likelihood already accounts for the differences between energy bands, so it would be unlikely for the thresholds to depend on energy band.

\subsection{Blind Survey Comparison between \wavlex and CSC Classifications}\label{subsec:survey}
As a final direct test of the efficacy of \wavlex, we compared the \wavlex likelihood classifications against those of the CSC pipeline for a common set of sources within the CLS. Unlike for the exceptionally deep LW observations, there is no obvious "ground truth" for the CLS sources, making it very challenging to confirm quantitatively whether the CSC or \wavlex is more complete and/or reliable for the CLS. We therefore conducted a blind survey in which members of this collaboration and other affiliated astrophysicists gave subjective scores -- based on visual evaluation -- to a list of CLS source candidates that were detected by both \wavlex and CSC. The ten survey respondents, all with experience in X-ray astronomy, each visually rated the same list of 100 source candidates on a scale from 1 to 5, where 1 is `very unlikely' and 5 is `very likely' to be a real source. Further details on how the survey results were analyzed are in Appendix \ref{sec:AD}. Visual evaluation may be subjective, but it is an established method for validating astronomical sources, as used by the \Chandra\ point-source catalog ChaMP \citep{Kim_2007}. Aggregated ratings from many observers have also been shown to agree well with expert classifications $\gtrsim90\%$ of the time in similar astronomical image recognition tasks like identifying galaxy morphology \citep{Willett_2013}. Because visual evaluation is subjective, we use the ratings only to compare the relative reliability of the two catalogs; the absolute reliability of \wavlex is set by the simulations of \S\ref{subsec:marxflux} and the ground-truth calibration of \S\ref{subsec:slices}. Mean ratings of MARX sources in Appendix~\ref{sec:AD} found similar reliability against flux and $\theta$ as \wavlex in Figure~\ref{fig:limit_heatmaps}.

\begin{figure}
    \centering
    \includegraphics[width=1\linewidth]{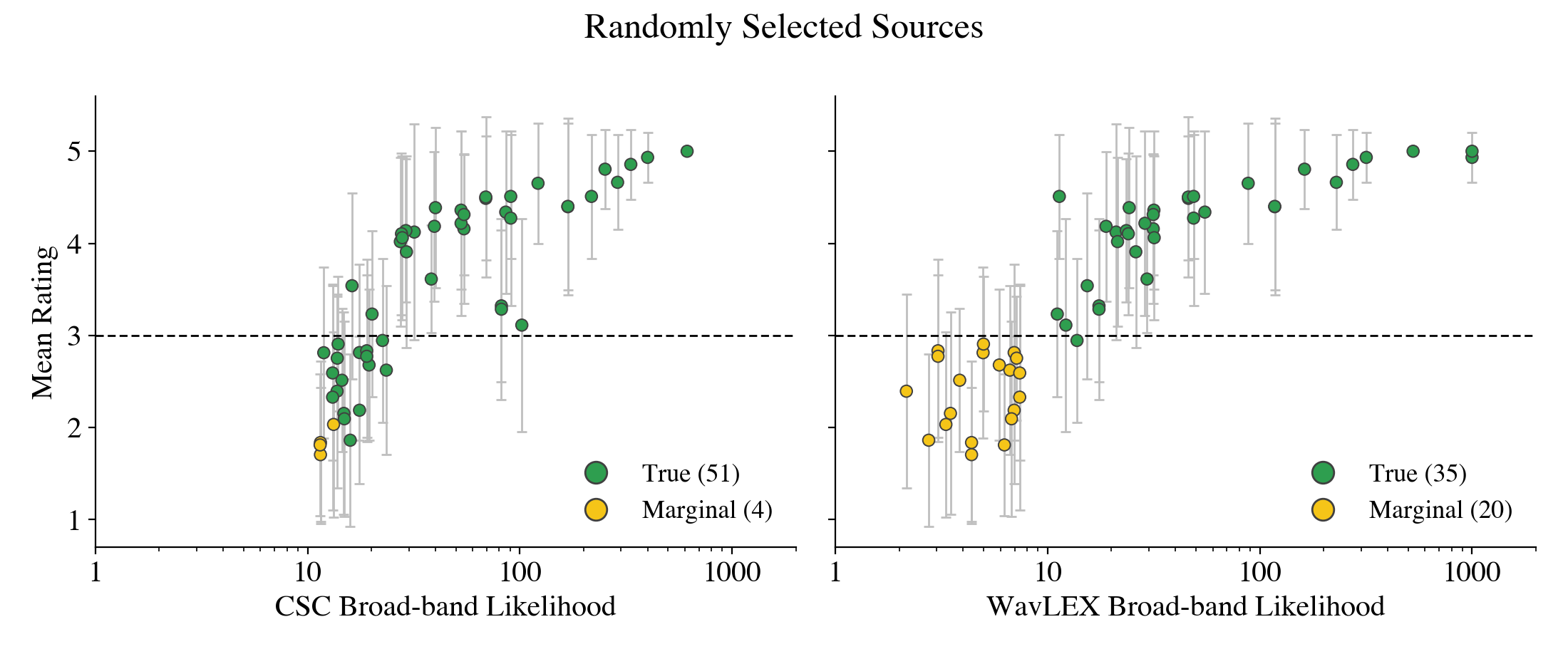}
    \caption{Mean ratings of 45 randomly selected CLS sources and 10 duplicates, spanning a range of likelihood, flux, and off-axis angle $\theta$, versus CSC likelihood (left) and \wavlex likelihood (right). Points are colored by their classification as True (green) or Marginal (yellow) by the CSC (left) and \wavlex (right). Error bars show the standard deviation of each source's ratings. The dashed horizontal line marks a mean rating of 3, where respondents considered the source equally likely to be real or spurious.}
    \label{fig:stamps_random}
\end{figure}

Figure~\ref{fig:stamps_random} shows the weighted mean survey ratings of the 45 randomly selected broad-band True and Marginal CLS sources as well as the 10 duplicate sources (discussed in Appendix \ref{sec:AD}). The CSC does not publish source candidates it believes to be `False' or `Unlikely,' so all CSC sources are Marginal at worst; to keep the survey blind, respondents were not told this. As expected, both the CSC and \wavlex likelihoods generally correlate positively with the mean ratings. However, 21 of the 55 sources received mean ratings $\leq3$, indicating that respondents considered them more likely to be spurious events than real sources. Of those 21 sources, the CSC classified $81\%$ as True and $19\%$ as Marginal, whereas all of the 20 Marginal \wavlex sources had mean ratings $\leq3$. Of the 51 "True" CSC sources, only $66.7\%$ had mean ratings $\geq3$, compared to $97.1\%$ of the 35 True \wavlex sources. 

Note that the CSC classifies sources using the highest likelihood across all energy bands. The broad-band likelihood is often the highest, but this is not guaranteed. If \wavlex sources were classified the same way, two of the 20 Marginal sources would have $\mathcal{L}>10$ in another band and become True, lowering the fraction of True \wavlex sources with mean ratings $\geq3$ from $97.1\%$ to $91.9\%$ (34 of 37), still above the $66.7\%$ for the CSC. The median standard deviation of the ratings for these sources is 0.88, incorporating weights discussed in Appendix \ref{sec:AD}.

\begin{figure}
    \centering
    \includegraphics[width=1\linewidth]{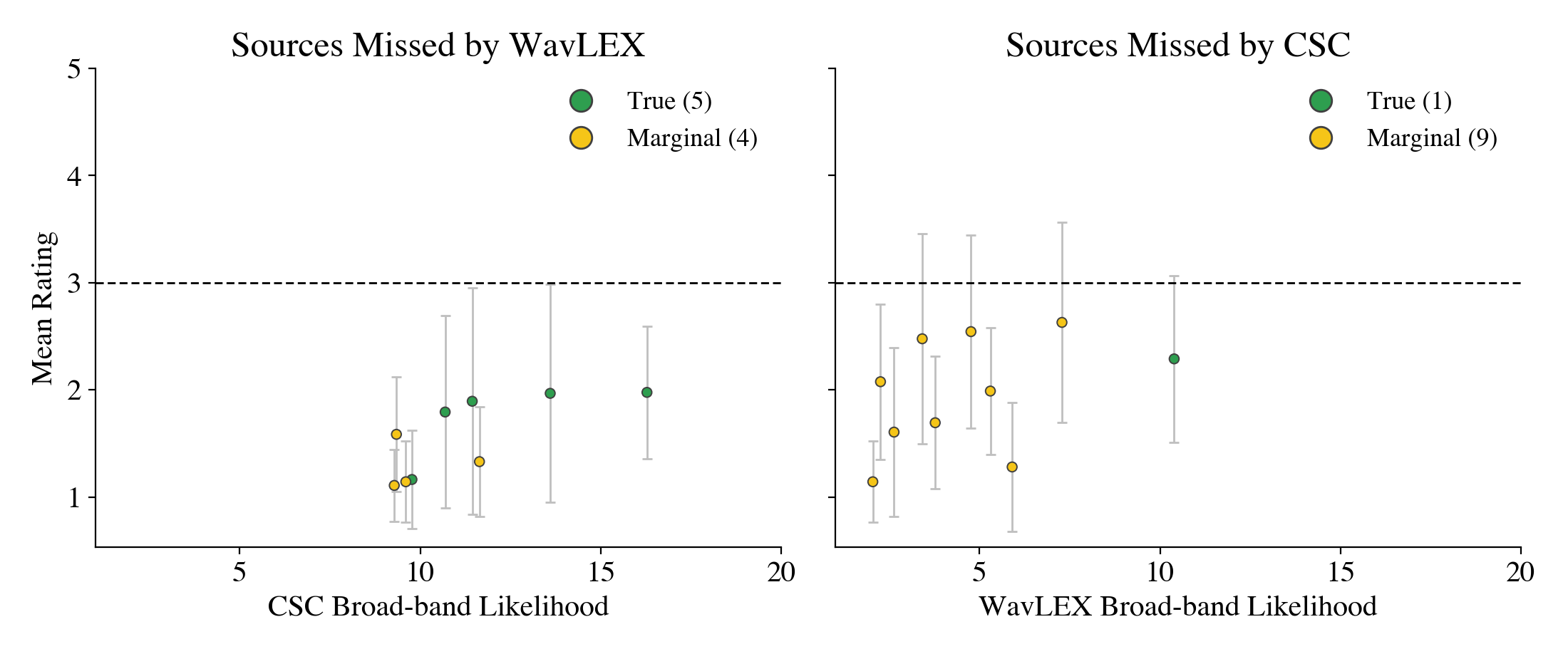}
    \caption{Mean ratings of 9 CLS source candidates missed by \wavlex versus their CSC likelihood (left), and of 10 source candidates missed by the CSC versus their \wavlex likelihood (right). Points are colored by their classification as True or Marginal by the CSC (left) and \wavlex (right). Error bars show the standard deviation of the ratings. The dashed horizontal line marks a mean rating of 3, where respondents considered the source equally likely to be real or spurious.}
    \label{fig:stamps_missed}
\end{figure}

One drawback of rating the likelihood of a set of sources that could all potentially be True is that false negatives are not accounted for. For this reason, the survey also included sets of sources missed by one of the two catalogs; the resulting ratings are shown in Figure~\ref{fig:stamps_missed}. All of the randomly selected source candidates missed by either the CSC or \wavlex had mean survey ratings $\leq3$. The nine CSC sources missed by \wavlex -- five of which were classified as True by the CSC -- had a mean rating with standard error of $1.52\pm0.11$. The ten \wavlex sources missed by the CSC -- one of which was classified as True by \wavlex -- had a mean rating of $1.97\pm0.17$. The two mean ratings differ by $0.45\pm0.20$ or $2.2\sigma$. Manual inspection showed that the source classified as True by \wavlex was a false detection caused by stray light from the brightest source in the CLS (photon flux $\sim(1.545\pm0.025)\times10^{-2}$~photons~cm$^{-2}$~s$^{-1}$), whereas the outlier candidate detected by the CSC ($\mathcal{L}=64.2$) showed no remarkable features but was excluded regardless. Future versions of \wavlex will be improved to recognize and exclude similar artifacts.

The survey also compared mean rating versus CSC and \wavlex likelihood for sources in the soft- and hard-band and simulated MARX sources. Ten sources shown twice during the survey received mean ratings differing by only $0.191\pm0.171$ on average, indicating that respondents answered consistently. This is discussed further in Appendix \ref{sec:AD}.

The main takeaways from this survey are:
\begin{itemize}
    \item Sources classified as True by \wavlex are rated as likely real (mean rating $\geq3$) in 34/35 ($\sim$97.1\%) of cases ($\sim$91.9\% when also including sources that are True in other energy bands), compared with 34/51 ($\sim$66.7\%) for the CSC.
    \item Sources detected only by \wavlex have a mean rating of $1.97\pm0.17$, compared to $1.52\pm0.11$ for sources detected only by the CSC, $2.2\sigma$ evidence that \wavlex may be more reliable.
    \item Respondents rate source candidates consistently (Figure~\ref{fig:duplicates}, Appendix \ref{sec:AD}), and these ratings are effective for making relative likelihood comparisons. 
\end{itemize}

\section{Results and Discussion}\label{sec:results}
Once \wavlex was fully developed and extensively tested, we used it to produce the final CLS catalog. \S\ref{subsec:catalog} describes the production of the catalog and provides basic information on the number of sources and their likelihood distributions. \S\ref{subsec:csc} compares these results to CSC~2.2 to evaluate whether \wavlex produces a reliable and complete catalog. Appendix~\ref{sec:AE} contains an excerpt of twelve sources from the completed catalog in Table~\ref{tab:excerpt} along with descriptions of included source properties. The full CLS catalog along with in depth analysis of source properties will be presented in future papers in this series. 

\subsection{Chandra Legacy Survey Source Catalog}\label{subsec:catalog}
The CLS consists of 350 \Chandra\ ACIS-I observations, each of which was processed individually with \wavlex in three main energy bands: soft (0.5--2~keV), broad (0.5--8~keV), and hard (2--8~keV). The hard-band catalog is most relevant to the study of the GC because extinction severely limits photon penetration below 2~keV; however, some of the detected hard-band sources are expected to be foreground or background sources. The soft-band catalog is expected to contain primarily foreground sources, while the broad-band catalog provides a more comprehensive list of sources with improved statistics. Additionally, a medium-band catalog (1.2--2~keV) was produced as a cross-check against the soft-band catalog; it is discussed in Appendix~\ref{sec:AC}.

To produce a single source list from the 350 CLS observations, each source was assigned a $2\sigma$ search radius from its centroid position uncertainty (with $s=1.21$; \S\ref{subsec:marxastro}). Two sources from separate observations were reported as a single source if the distance between their positions was less than $\sqrt{r_1^2+r_2^2}$, where $r_1$ and $r_2$ are their $2\sigma$ search radii, with catalog parameters taken from the observation with the highest likelihood. A group of three or more sources was merged only if every pair satisfied this criterion. The final catalog includes only sources with a True or Marginal classification (\S\ref{subsec:mle}).

Across the 350 observations, \wavdetect produced source candidates in each of the three energy bands. In the broad band, for example, \wavdetect detected 10,311 candidates in total, of which 9886 remain after removing duplicates from overlapping observations. After removing duplicates in each band, these candidates were classified as True, Marginal, Unlikely, or Too Faint (\S\ref{subsec:mle}); only True and Marginal sources are included in the catalog. Table~\ref{tab:catalog_amounts} gives the number of candidates, True and Marginal sources, and sources detected in only one band for each energy band and for the merged catalog.

\setlength{\tabcolsep}{6pt}
\begin{deluxetable}{lcccccc}
\tablecaption{Sources in the CLS catalog by energy band. Detected: unique \wavdetect source candidates after removing duplicates from overlapping observations. Catalog total: True + Marginal sources. Band only: sources detected in only one of the three bands. For the merged catalog, Detected is the sum of the three bands (only True and Marginal sources were crossmatched across bands), and the last column instead gives the number of sources detected in all three bands. \label{tab:catalog_amounts}}
\tablehead{
\colhead{Band} & \colhead{Energy (keV)} & \colhead{Detected} & \colhead{True} & \colhead{Marginal} & \colhead{Catalog total} & \colhead{Band only}
}
\startdata
Soft  & 0.5--2 & 4505 & 545  & 819  & 1364 & 491  \\
Hard  & 2--8   & 6970 & 1269 & 1980 & 3249 & 616  \\
Broad & 0.5--8 & 9886 & 1688 & 2670 & 4358 & 1245 \\
\hline
Merged & 0.5--8 & 21361 & 1989 & 3472 & 5461 & 339 \\
\enddata
\end{deluxetable}
\setlength{\tabcolsep}{15pt}

\begin{figure}
    \centering
    \includegraphics[width=0.8\linewidth]{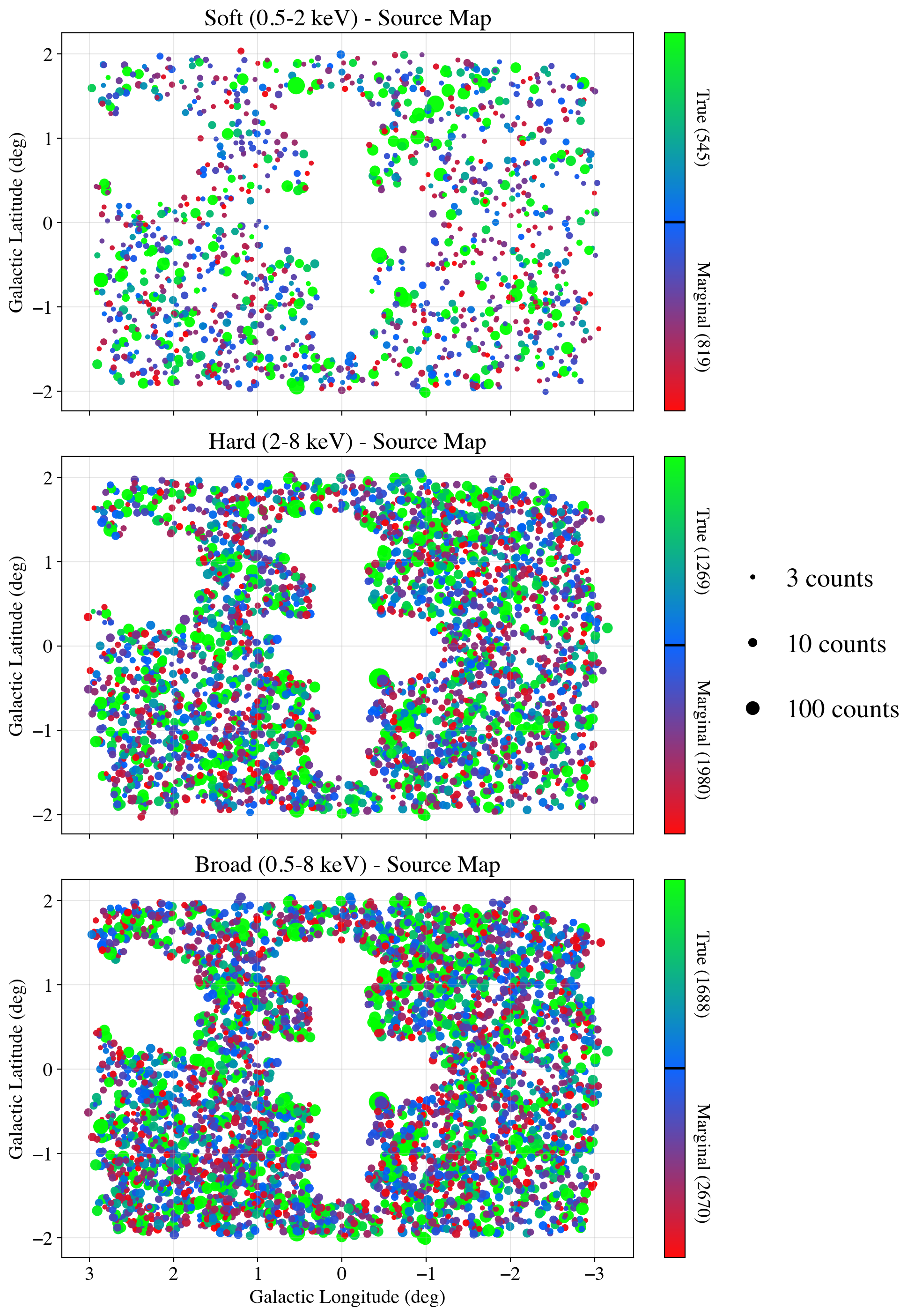}
    \caption{Source maps of the CLS catalog in each energy band: soft (0.5--2 keV; top), hard (2--8 keV; middle), and broad (0.5--8 keV; bottom). The color of each source indicates its likelihood: $\mathcal{L}=2.5$ is red, $\mathcal{L}=10$ is blue, and the maximum $\mathcal{L}$ is green. The size of each source scales with the number of counts, as indicated by the legend on the right. The number of True and Marginal sources in each source list is also indicated on the color bar.}
    \label{fig:sourcemaps}
\end{figure}

Assuming the soft catalog is the complete catalog of foreground sources, the foreground source ratio is $N_{\mathrm{soft}}/N_{\mathrm{all}}=24.98\%$, implying a moderate proportion of detected foreground sources. However, this should not be taken as a reflection of the foreground vs GC source density, as there are observational biases, such as differences in effective area (Figure~\ref{fig:effarea}) and the lower $N_{\mathrm{H}}$ at greater $|b|$ (Figure~\ref{nhcontor}). The source maps for each of the three main energy bands are shown in Figure~\ref{fig:sourcemaps}.

While each energy band can serve as its own independent catalog, we also produced a master catalog containing all sources. Using the same method of crossmatching $2\sigma$ uncertainty radii added in quadrature, we found 5461 unique sources across all catalogs, with 1989 True sources and 3472 Marginal sources. 

\begin{figure}
    \centering
    \includegraphics[width=1\linewidth]{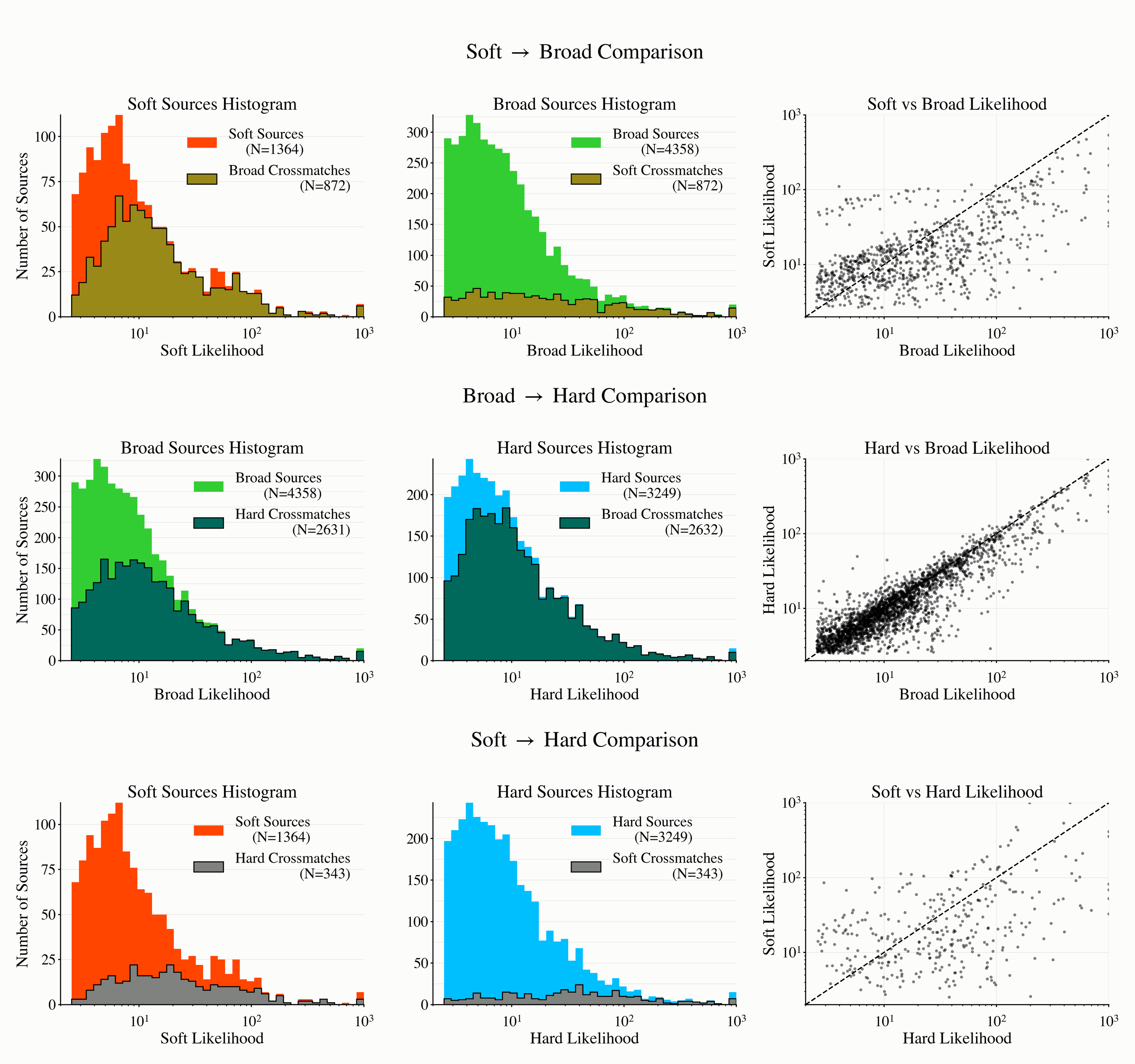}
    \caption{The left and middle columns show the distributions of likelihoods $\mathcal{L}$ for sources in each energy band, and the right column shows scatter plots comparing the likelihoods of crossmatched sources in two energy bands. Each row pairs two energy bands: soft and broad (top), broad and hard (middle), and soft and hard (bottom). In each histogram, the full catalog of one band is overlaid with the subset of its sources crossmatched to the other band using the band-specific likelihood for the x-axis. The energy bands are colored blue (for hard), red (for soft), and green (for broad). The legends give the group shown in each histogram and the number of sources in that group. There are 40 logarithmically spaced bins for $2.5\leq\mathcal{L}\leq1000$.}
    \label{fig:crossmatchhist}
\end{figure}

The completion of each energy band's source list relative to the others is also shown as a function of likelihood in Figure~\ref{fig:crossmatchhist}. The most populous sets of crossmatched sources are 2632 hard--broad sources, followed by 872 soft--broad sources, and 343 soft--hard sources. The likelihood bins with the highest proportion of sources in Figure~\ref{fig:crossmatchhist} are $3.92<\mathcal{L}\leq4.55$ for the broad-band, $3.92<\mathcal{L}\leq4.55$ for the hard-band, and $6.16<\mathcal{L}\leq7.16$ for the soft-band. Given that flux correlates with likelihood and that detectability as a function of flux was already explored in \S\ref{subsec:marxflux}, similar histograms of flux could be used to estimate the \emph{intrinsic} flux distribution for sources in the CLS. These and other analyses will be presented in subsequent papers in this series.

\subsection{Comparison with the CSC}\label{subsec:csc}
CSC~2.2 was released on October 5, 2026, marking the latest update to the CSC \citep{Evans_2024_ApJS}. It covers all \Chandra\ observations that were publicly available by the end of 2023, including the CLS, so it is worth comparing the 3619 CLS sources in CSC~2.2 to the 5461 CLS sources detected by \wavlex. \wavlex and the CSC pipeline share the same core methods: \wavdetect for source detection and an MLE algorithm for source validation. The pipelines diverge in that the CSC is designed to analyze all \Chandra\ observations over the entire sky, whereas \wavlex is optimized for source detection in low-exposure, strong-background observations like the CLS. In \S\ref{subsec:survey}, we compared the CSC and \wavlex with a survey visually rating individual sources; this section instead examines the CLS catalog as a whole, with some extrapolations from \S\ref{subsec:survey}.

The master catalog of the CSC contains 3619 sources for the CLS ObsIDs 23884--24233. The CSC differs from \wavlex in that, instead of publishing catalogs for each energy band, it publishes a master catalog with dependent source characteristics such as likelihood included for each energy band. As such, the master catalogs from \wavlex and CSC are best suited for direct comparison. In order to make relative comparisons with the CSC for specific energy bands, we filtered for CSC sources in the master catalog with a strictly positive band-specific likelihood, which avoids any assumptions about the CSC likelihood thresholds. We will keep the convention of referring to these filtered CSC source lists as band catalogs even though the CSC does not use these terms. Note that the CSC broad and hard bands extend only to 7~keV, compared to 8~keV for \wavlex, although the \Chandra\ ACIS-I effective area is significantly diminished above 7~keV (Figure~\ref{fig:effarea}). 

The CSC broad-band catalog contains 3599 sources, compared to 4358 from \wavlex, and the CSC hard-band catalog contains 2854 sources, vs. 3249 from \wavlex. The CSC soft and medium bands together span the same energy range as the \wavlex soft band, and the combined CSC soft and medium catalogs contain 670 sources, compared to 1364 from \wavlex. The \wavlex hard-band and broad-band catalogs contain $13.8\%$ and $21.1\%$ more sources than the CSC, respectively. There are additional factors to consider when comparing the soft and medium bands of the CSC to \wavlex, so that comparison is discussed in Appendix~\ref{sec:AC}.

\begin{figure}
    \centering
    \includegraphics[width=1\linewidth]{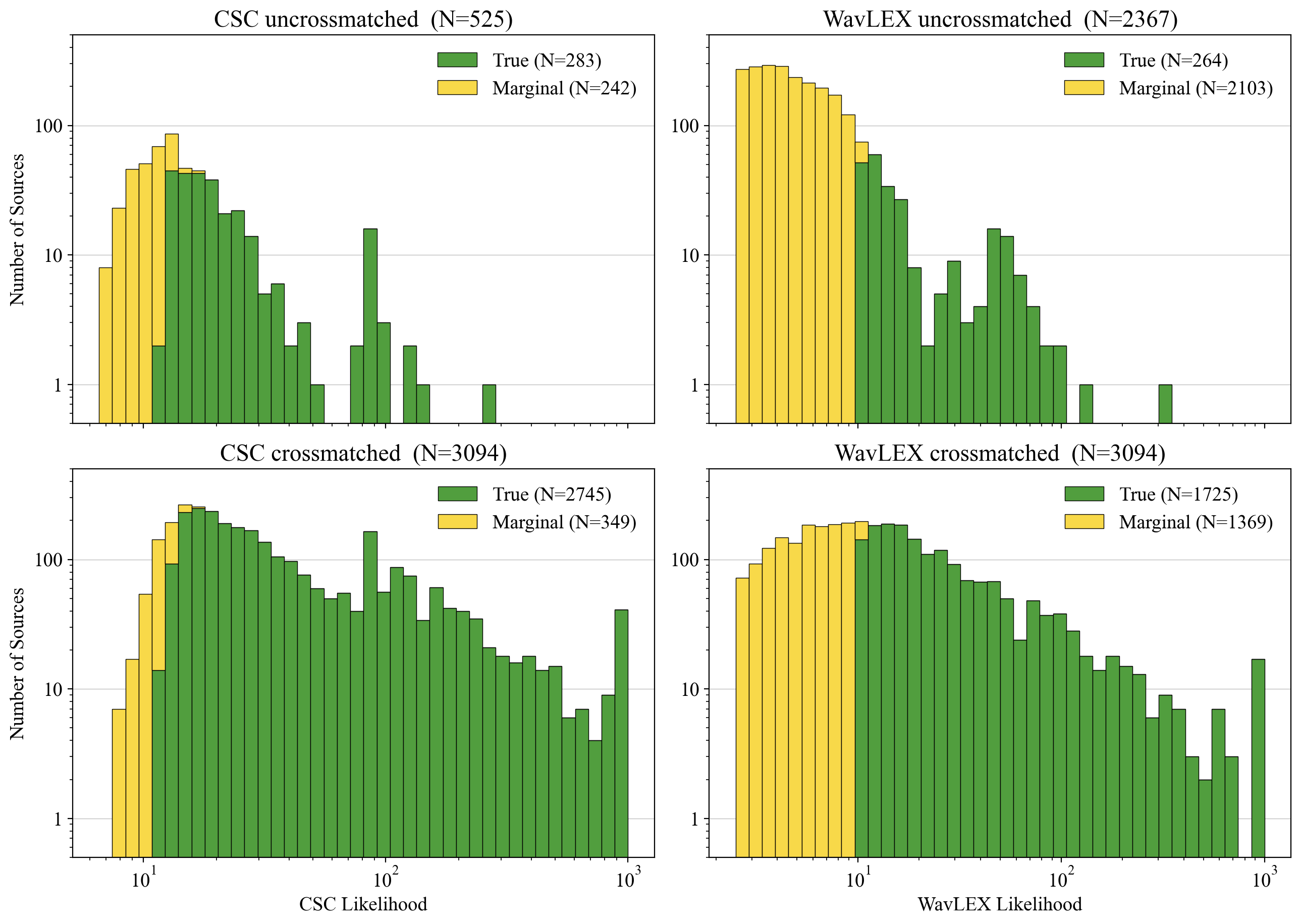}
    \caption{Likelihood histograms of the CLS catalogs from the CSC and \wavlex. Histograms on the left show CSC likelihood, and histograms on the right show \wavlex likelihood. The bottom row shows the set of crossmatched sources found in both the CSC and \wavlex, and the top row shows CSC sources missed by \wavlex (top left) and \wavlex sources missed by the CSC (top right). Legends indicate the number of True and Marginal sources, and the titles show the total number of sources in each panel. Each histogram has 40 logarithmically spaced bins.}
    \label{fig:csc_wavlex_hist}
\end{figure}

To compare the completeness of the two catalogs, we crossmatched the CSC and \wavlex sources using the same method as in \S\ref{subsec:slices}. Figure~\ref{fig:csc_wavlex_hist} shows the likelihood histograms of the CSC-only, \wavlex-only, and crossmatched sets of sources.

The total proportion of sources detected by the CSC marked as True is not published, but as shown in Figure~\ref{fig:csc_wavlex_hist} the CSC classifies a higher proportion of its True + Marginal sources as True ($83.7\%$), compared to $36.4\%$ for \wavlex. However, \wavlex and the CSC pipeline use different thresholds for classifying a source as True. For the same set of 3094 crossmatched sources, \wavlex classified $55.8\%$ as True (i.e., $\geq2\sigma$ confidence), while the CSC classified $88.7\%$ as True. More informative than these differing classification thresholds are the 525 source candidates missed by \wavlex, of which $53.9\%$ are classified as True by the CSC, and the 2367 source candidates missed by the CSC, of which $11.2\%$ are classified as True by \wavlex.

Referring to Figure~\ref{fig:stamps_missed} from \S\ref{subsec:survey}, 10\% of the uncrossmatched \wavlex source candidates in the survey were classified by the catalog as True (although that one True source was an outlier), as were 55.6\% of the uncrossmatched CSC source candidates. These fractions are comparable to those of the full sets of uncrossmatched sources. The mean rating of the uncrossmatched \wavlex source candidates in Figure~\ref{fig:stamps_missed} was $1.97\pm0.17$, compared to $1.52\pm0.11$ for the uncrossmatched CSC source candidates. Because these ratings have not been calibrated against an undisputed set of real sources, as the likelihoods were in \S\ref{subsec:slices}, they can only be used for relative comparisons. If the surveyed sources are comparable to the full sets of uncrossmatched sources, then the 2367 uncrossmatched \wavlex sources may well have a higher fraction of real sources than the 525 uncrossmatched CSC sources. This suggests that the \wavlex CLS catalog is far more complete and plausibly more reliable than the CSC~2.2 catalog of the same strong-background, low-exposure observations.

\section{Summary \& Conclusions}\label{sec:concl}
This paper is the first in a series that will use \Chandra\ observations of the central $4^{\circ} \times 6^{\circ}$ of the Galaxy, known as the \Chandra\ Legacy Survey (CLS), to systematically study and classify the point source populations of the GC. The CLS consists of relatively low-exposure observations ($\sim 5$~ks) taken within the strong-background GC, which are challenging conditions for source detection. This paper therefore focuses on the development of \wavlex -- an automated source detection and validation tool specialized for low-exposure, strong-background \Chandra\ observations -- and on how \wavlex was used to produce a CLS source catalog. Our main results are:

\begin{itemize}
    \item \textbf{A New Source Detection Tool (\S\ref{sec:wvlx}):} \wavlex automatically processes a list of ObsIDs into a source list, using \wavdetect to detect source candidates; it then derives a maximum likelihood estimate $\mathcal{L}$ that parametrizes the probability that a detected source is real.
    \item \textbf{Testing and calibration (\S\ref{sec:tests}):} MARX simulations show a flux sensitivity of $(1$--$20)\times10^{-14}$ erg cm$^{-2}$ s$^{-1}$ (depending on the off-axis angle $\theta$) and a median $2\sigma$ position uncertainty of $0.94''$ for the adopted PSF-core centroid method. Running \wavlex on $5$~ks slices of Limiting Window observations allows us to set the likelihood thresholds at $\mathcal{L}=2.5$ (corresponding to $\sim1\sigma$) for ``Marginal'' sources and $\mathcal{L}=10$ ($\sim2\sigma$) for ``True'' sources. In the blind survey, 91.9--97.1\% of sources classified as ``True'' by \wavlex were rated as likely to be real, compared to 66.7\% of sources classified as ``True'' by the CSC; and sources detected only by \wavlex had a weighted mean survey rating $2.2\sigma$ higher ($1.97\pm0.17$ out of $5$) than those detected only by the CSC ($1.52\pm0.11$ out of $5$), suggesting that \wavlex is likely more reliable than the CSC within the CLS region.
    \item \textbf{CLS Catalog (\S\ref{sec:results}):} We used \wavlex to produce a catalog of 5461 CLS point sources (1989 True, 3472 Marginal), with band-specific catalogs in the soft (0.5--2~keV; 1364 sources), hard (2--8~keV; 3249), and broad (0.5--8~keV; 4358) bands (Table~\ref{tab:catalog_amounts}). This includes 2367 sources that were not detected by the \Chandra\ Source Catalog, all with $>1\sigma$ confidence and reliability comparable to CSC sources.
\end{itemize}

The open-source \wavlex code is currently available on GitHub for use on any public \Chandra\ observations. The second paper in this series will present the analysis and scientific implications of the \wavlex catalog of the CLS.

\vspace{1cm}

\begin{acknowledgments}
This research has made use of data obtained from the Chandra Source Catalog and Chandra Data Archive, provided by the Chandra X-ray Center (CXC)\dataset[10.25574/cdc.750]{10.25574/cdc.750}. We thank the Chandra X-ray Center (CXC) for its high-resolution data, thorough documentation, and for hosting the Chandra Dynamic Universe conference, where this work was presented. We thank Drs. Jaegeun Park, Gabriel Bridges, Samaresh Mondal, and Giulia Brunelli for completing the blind survey described in \S\ref{subsec:survey}; the Center for Astrophysics | Harvard \& Smithsonian for hosting the first author in summer 2025; and Drs. Patrick Slane, Tom Maccarone, Peter Jonker, and David Kipping for helpful discussions and comments. This research has made use of data obtained from the Chandra Data Archive and software provided by the CXC in the software package CIAO. Shifra Mandel acknowledges support by the National Science Foundation Graduate Research Fellowship under Grant No. DGE 2036197 and the Columbia University Provost Fellows Program.  
\end{acknowledgments}

\facilities{CXO}
\software{CIAO v4.17 \citep{2026ApJ...1005..116F}, CALDB v4.12.4, Sherpa \citep{2024ApJS..274...43S} v4.17, wavdetect \citep{Freeman2002}, MARX v5.5.3 \citep{Wise2013MARX}, Astropy v8.0 \citep{astropy:2018}}

\appendix 

\section{Excerpt of the CLS Catalog}\label{sec:AE}

Table~\ref{tab:excerpt} lists twelve sources from the CLS catalog, spanning a range of likelihoods, energy bands, and off-axis angles, to illustrate its format and content; the full catalog will be released with the next paper in this series. Source names are IAU-style designations derived from the truncated centroid coordinates. \emph{Energy Bands} lists every band in which a source was detected, and the remaining columns describe the detection in the \emph{Best Band}, the one with the highest likelihood. The likelihood $\mathcal{L}$ and the True/Marginal classification are defined in \S\ref{subsec:mle}. The \wavdetect positions are described in \S\ref{subsec:wavdet} and the centroid positions and $2\sigma$ uncertainties in \S\ref{subsec:astrometry} with additional uncertainty $s$ included. The effective exposure $\bar{E}_{\rm eff}$, the line-of-sight column densities $N_{\rm HI}$ and $N_{\rm H_2}$, and the photon flux $F$ with its lower and upper $2\sigma$ offsets are described in \S\ref{subsec:additions}.

\movetabledown=1 in
\begin{splitdeluxetable*}{lcccccccccccccBlcccccccccc}
\rotate
\tabletypesize{\footnotesize}
\setlength{\tabcolsep}{3pt}
\tablecaption{Twelve source excerpt of the CLS Catalog. \label{tab:excerpt}}
\tablehead{
\colhead{Name} & \colhead{Energy} & \colhead{Best} & \colhead{ObsIDs} & \colhead{$\mathcal{L}$} & \colhead{Class} & \multicolumn{4}{c}{Centroid} & \multicolumn{4}{c}{\texttt{wavdetect}} &
\colhead{Name} & \colhead{$\theta$} & \colhead{Counts} & \colhead{96\% PSF} & \colhead{$\bar{E}_{\rm eff}$} & \colhead{Background} & \colhead{$N_{\rm HI}$} & \colhead{$N_{\rm H_2}$} & \multicolumn{3}{c}{Photon flux} \\
\colhead{} & \colhead{Bands} & \colhead{Band} & \colhead{} & \colhead{} & \colhead{} & \colhead{RA} & \colhead{Dec} & \colhead{$2\sigma_{\rm RA}$} & \colhead{$2\sigma_{\rm Dec}$} & \colhead{RA} & \colhead{Dec} & \colhead{$2\sigma_{\rm RA}$} & \colhead{$2\sigma_{\rm Dec}$} &
\colhead{} & \colhead{(arcmin)} & \colhead{} & \colhead{Size} & \colhead{(cm$^2$~s)} & \colhead{Rate} & \colhead{(cm$^{-2}$)} & \colhead{(cm$^{-2}$)} & \multicolumn{3}{c}{(photons~cm$^{-2}$~s$^{-1}$)} \\
\colhead{} & \colhead{} & \colhead{} & \colhead{} & \colhead{} & \colhead{} & \colhead{(deg)} & \colhead{(deg)} & \colhead{(arcsec)} & \colhead{(arcsec)} & \colhead{(deg)} & \colhead{(deg)} & \colhead{(arcsec)} & \colhead{(arcsec)} &
\colhead{} & \colhead{} & \colhead{} & \colhead{(pix)} & \colhead{} & \colhead{(ct~pix$^{-1}$)} & \colhead{} & \colhead{} & \colhead{$F$} & \colhead{$-2\sigma\ F$} & \colhead{$+2\sigma\ F$}
}
\startdata
CXO~J173836$-$281344 & soft+broad & soft & 24048 & 15.77 & True & 264.652798 & $-28.228919$ & 0.29 & 0.43 & 264.652798 & $-28.228919$ & 0.24 & 0.41 & CXO~J173836.6$-$281344.1 & 2.65 & 4 & 9 & 482543.3 & 0.0005112 & $5.63\times10^{21}$ & $3.56\times10^{20}$ & $1.035\times10^{-5}$ & $6.156\times10^{-6}$ & $1.617\times10^{-5}$ \\
CXO~J174024$-$285652 & broad & broad & 24022 & 50.92 & True & 265.100267 & $-28.947789$ & 0.56 & 0.65 & 265.100708 & $-28.947474$ & 0.72 & 0.56 & CXO~J174024.0$-$285652.0 & 5.33 & 23 & 75 & 1507419.2 & 0.009974 & $8.94\times10^{21}$ & $3.54\times10^{21}$ & $1.895\times10^{-5}$ & $6.321\times10^{-6}$ & $9.546\times10^{-6}$ \\
CXO~J174037$-$265046 & soft+broad & broad & 24007 & 6.69 & Marginal & 265.157378 & $-26.846137$ & 2.83 & 1.41 & 265.157621 & $-26.846135$ & 3.77 & 1.72 & CXO~J174037.7$-$265046.0 & 10.38 & 13 & 952 & 1180140.1 & 0.09791 & $4.28\times10^{21}$ & $5.25\times10^{20}$ & $1.245\times10^{-5}$ & $5.663\times10^{-6}$ & $9.777\times10^{-6}$ \\
CXO~J174248$-$262444 & hard & hard & 24052 & 3.02 & Marginal & 265.703171 & $-26.412268$ & 0.38 & 0.59 & 265.703171 & $-26.412268$ & 0.30 & 0.46 & CXO~J174248.7$-$262444.1 & 4.50 & 3 & 51 & 1828355.5 & 0.004680 & $4.59\times10^{21}$ & $3.39\times10^{20}$ & $2.006\times10^{-6}$ & $1.306\times10^{-6}$ & $3.943\times10^{-6}$ \\
CXO~J174425$-$310854 & broad & broad & 24176 & 13.50 & True & 266.108216 & $-31.148584$ & 1.52 & 0.53 & 266.108275 & $-31.148702$ & 1.57 & 0.92 & CXO~J174425.9$-$310854.9 & 5.73 & 5 & 90 & 1081574.4 & 0.004751 & $9.34\times10^{21}$ & $2.89\times10^{21}$ & $5.652\times10^{-6}$ & $3.234\times10^{-6}$ & $7.707\times10^{-6}$ \\
CXO~J174426$-$321012 & hard & hard & 23975 & 32.97 & True & 266.108600 & $-32.170179$ & 0.80 & 0.77 & 266.109090 & $-32.170098$ & 1.27 & 0.72 & CXO~J174426.0$-$321012.6 & 7.30 & 16 & 251 & 1451938.8 & 0.01608 & $6.86\times10^{21}$ & $1.25\times10^{21}$ & $1.358\times10^{-5}$ & $5.249\times10^{-6}$ & $8.595\times10^{-6}$ \\
CXO~J173931$-$294311 & soft+hard & hard & 24090 & 13.87 & True & 264.880140 & $-29.719986$ & 1.05 & 1.59 & 264.880229 & $-29.719708$ & 1.60 & 1.71 & CXO~J173931.2$-$294311.9 & 7.77 & 11 & 327 & 1188521.3 & 0.03264 & $1.08\times10^{22}$ & $2.62\times10^{21}$ & $1.123\times10^{-5}$ & $5.048\times10^{-6}$ & $9.131\times10^{-6}$ \\
CXO~J175133$-$285918 & broad+hard & hard & 23943 & 5.91 & Marginal & 267.890009 & $-28.988552$ & 1.43 & 1.68 & 267.890519 & $-28.988665$ & 2.30 & 0.93 & CXO~J175133.6$-$285918.7 & 8.22 & 8 & 394 & 1573284.9 & 0.04315 & $7.2\times10^{21}$ & $1.02\times10^{21}$ & $6.001\times10^{-6}$ & $3.086\times10^{-6}$ & $6.168\times10^{-6}$ \\
CXO~J175233$-$293945 & broad+hard & broad & 23939 & 239.16 & True & 268.138624 & $-29.662679$ & 0.18 & 0.21 & 268.138564 & $-29.662693$ & 0.26 & 0.28 & CXO~J175233.2$-$293945.6 & 4.56 & 52 & 45 & 1135344.7 & 0.007768 & $5.12\times10^{21}$ & $1.15\times10^{20}$ & $5.716\times10^{-5}$ & $1.354\times10^{-5}$ & $1.783\times10^{-5}$ \\
CXO~J174627$-$295821 & soft+broad+hard & hard & 23897 & 85.61 & True & 266.615789 & $-29.972769$ & 0.55 & 1.37 & 266.616041 & $-29.972944$ & 0.98 & 1.50 & CXO~J174627.7$-$295821.9 & 9.06 & 33 & 584 & 1554360.5 & 0.06841 & $1.17\times10^{22}$ & $7.19\times10^{21}$ & $2.482\times10^{-5}$ & $7.603\times10^{-6}$ & $1.073\times10^{-5}$ \\
CXO~J174715$-$295801 & soft & soft & 23891 & 172.91 & True & 266.814355 & $-29.967194$ & 1.00 & 0.64 & 266.814335 & $-29.967002$ & 1.11 & 0.91 & CXO~J174715.4$-$295801.8 & 10.61 & 82 & 928 & 213176.0 & 0.1244 & $1.03\times10^{22}$ & $6.21\times10^{21}$ & $4.783\times10^{-4}$ & $9.318\times10^{-5}$ & $1.160\times10^{-4}$ \\
CXO~J175551$-$283213 & soft+broad+hard & broad & 23930 & 417.38 & True & 268.965009 & $-28.537077$ & 0.49 & 0.33 & 268.964939 & $-28.537023$ & 0.55 & 0.38 & CXO~J175551.6$-$283213.4 & 7.54 & 106 & 288 & 1169777.5 & 0.03231 & $4.07\times10^{21}$ & $4.98\times10^{20}$ & $1.128\times10^{-4}$ & $1.957\times10^{-5}$ & $2.373\times10^{-5}$ \\
\enddata
\end{splitdeluxetable*}
\section{Edge-Case Tests of \wavlex}\label{sec:AB}
 
Some observing conditions, mostly unusual ones, can interfere with the detection and evaluation of a source by \wavlex. Many of these circumstances are improbable enough to be ignored, with negligible impact on the CLS or on other observations that \wavlex may be used to process. However, we considered two cases worth investigating: the effects of source crowding and of dust scattering on \wavlex.

\subsection{Source crowding}\label{subsec:crowding}
 
Source crowding occurs when two or more sources are close enough to each other that an algorithm such as \wavlex may attribute photons from one source to another or, in severe cases, identify multiple sources as a single source. This makes extracted source properties much less reliable.  We therefore investigate the conditions under which source parameters are significantly affected. We used MARX to simulate pairs of crowded sources and evaluated how their \wavlex detection is affected depending on their separation, off-axis angle $\theta$, and flux, using the same flux for each source in a pair. We adopt the ranges $Flux \in \{3.3\times10^{-14},\,1\times10^{-13},\,3.3\times10^{-13}\}~\mathrm{erg\,cm^{-2}\,s^{-1}}$, $\theta \in \{2,\,6,\,10\}~\mathrm{arcmin}$, and $\text{separation} \in \{1,\,2,\,4\}~\mathrm{arcsec}$. The results are shown in Table~\ref{tab:crowding}.
 
\begin{deluxetable}{lccc}
\tablecaption{\wavlex results for crowded pairs of MARX sources. The Detection column shows whether a pair of sources was marked as two separate sources or as a single combined source by \wavdetect. The control values were ($\theta=6^{\prime}$, flux $=10^{-13}$~erg~cm$^{-2}$~s$^{-1}$, separation $=2^{\prime\prime}$). \label{tab:crowding}}
\tablehead{
\colhead{Test variable} & \colhead{Detection} & \colhead{Likelihood} & \colhead{Classification}
}
\startdata
Control & Combined & 18 & True \\
$\theta=2^{\prime}$ & Separate & 26 & True \\
$\theta=10^{\prime}$ & Combined & 5 & Marginal \\
Flux $=3.3\times10^{-14}$~erg~cm$^{-2}$~s$^{-1}$ & Combined & 4 & Marginal \\
Flux $=3.3\times10^{-13}$~erg~cm$^{-2}$~s$^{-1}$ & Combined & 49 & True \\
Separation $=1^{\prime\prime}$ & Combined & 23 & True \\
Separation $=4^{\prime\prime}$ & Separate & 11 & True \\
\enddata
\end{deluxetable}
 
Table~\ref{tab:crowding} shows that the sources were marked as distinct only for an off-axis angle of $2^{\prime}$ or a separation of $4^{\prime\prime}$, whereas flux had no impact on detection in these trials (possibly because of the range simulated). These results are intuitive: at greater separation, sources are easier to resolve; and at smaller off-axis angles, the respective PSFs of crowded sources are smaller and easier to distinguish. Among the small-, medium-, and large-separation pairs, the large-separation pair has the lowest individual likelihoods because its sources are evaluated individually, each with half of the combined flux. The small-separation pair has the highest likelihood, possibly because in combination, it better approximates a real PSF shape than the medium-separation pair. To put the effect of source crowding on the CLS in perspective, 10,311 source candidates were detected with \wavdetect over 350 observations before applying any filters. The ACIS-I detector is $\sim 16.9^{\prime}\times16.9^{\prime}$; if we assume that sources are randomly distributed, the average off-axis angle is $\sim 6.5^{\prime}$, similar to the $6^{\prime}$ where sources were combined at $<4^{\prime\prime}$. If we also assume that $4^{\prime\prime}$ is the minimum separation required to avoid crowding for the average source (for context, the average distance to the nearest neighbor in the CLS is $\sim154''$), then the probability of a pair of crowded sources being marked as a single source in the CLS catalog is $\sim 0.14\%$, which is well below the acceptable limit.

\subsection{Dust scattering}\label{subsec:dust}

\begin{figure}
    \centering
    \includegraphics[width=0.9\linewidth]{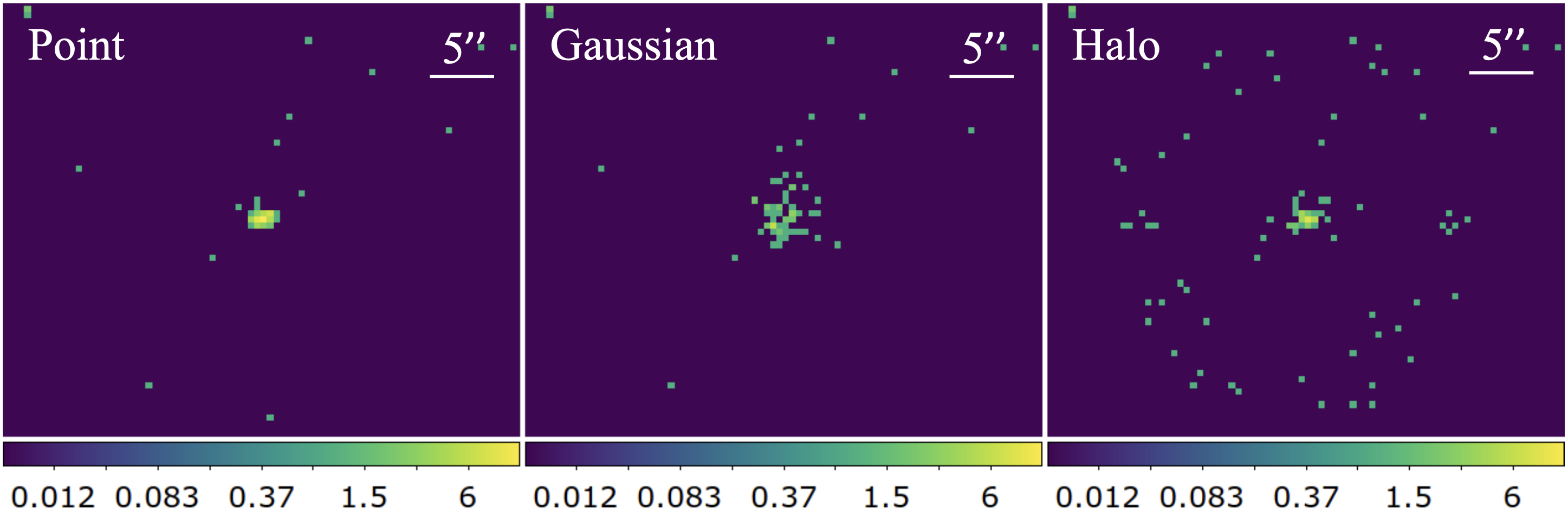}
    \caption{MARX Images of the same point source in three scattering configurations, each with flux $10^{-13}$~erg~cm$^{-2}$~s$^{-1}$ and off-axis angle $\theta=2^{\prime}$. Left: point source without dust scattering. Center: point source with dust scattering, simulated by a Gaussian of $\sigma=2^{\prime\prime}$.  Right: point source with dust scattering "halo" at a radius $r=10^{\prime\prime}$--$15^{\prime\prime}$. A $5''$ ruler is shown in the upper right for scale.}
    \label{fig:dustimage}
\end{figure}

The field of view of the CLS includes regions of considerably high $N_{\mathrm{H}}$, up to $10^{23}$~cm$^{-2}$. This can indicate a high dust concentration along the line of sight, which can distort a source image before it reaches the \Chandra\ optics \citep{2017MNRAS.468.2532J}. Dust scattering depends on many variables, including grain size, grain composition, distance to the dust, and dust column density, so modeling it precisely for individual sources is beyond the scope of this work. Its common signatures, however, are well established: point sources appear extended (approximated here by a Gaussian) or are surrounded by a diffuse halo \citep{1991ApJ...376..490M}. Rather than deriving parameters from a specific dust model, we adopted generalized, representative values for these two effects to test how sensitive \wavlex is to them. Using MARX, we injected simulated sources into observations with all variables held constant except the source type: (1) a ``normal'' point source; (2) a scattered source modeled as a Gaussian with $\sigma=2^{\prime\prime}$, a modest broadening comparable to the \Chandra\ PSF scale; and (3) a point source with a dust-scattering halo at $r=10^{\prime\prime}$--$15^{\prime\prime}$ from the source centroid, where the halo flux equals the central point-source flux. The total (center + halo) flux is twice the listed value. Images of one such set of sources are shown in Figure~\ref{fig:dustimage}. This was repeated for all combinations of point-source $Flux \in \{3.3\times10^{-14},\,1\times10^{-13},\,3.3\times10^{-13}\}~\mathrm{erg\,cm^{-2}\,s^{-1}}$ and $\theta \in \{2,\,5,\,8\}~\mathrm{arcmin}$. \wavlex was then run on all simulated sources, with the results shown in Table~\ref{tab:dust}.
 
\begin{deluxetable}{llccc}
\tablecaption{\wavlex classification of simulated MARX sources \label{tab:dust}}
\tablehead{
\colhead{Source Type} & \colhead{} & \multicolumn{3}{c}{Flux (erg\,cm$^{-2}$\,s$^{-1}$)} \\
\colhead{} & \colhead{} & \colhead{$3.30\times10^{-14}$} & \colhead{$1.00\times10^{-13}$} & \colhead{$3.30\times10^{-13}$}} 
\startdata
\noalign{\vskip 6pt}
$\ \ \ \ \ \ \ $Point sources (no scattering) & & 
& & \\
\noalign{\vskip 6pt}\hline\noalign{\vskip 6pt}
                & $2^{\prime}$ & True         & True     & True \\
Off-axis angle (arcmin): & $5^{\prime}$ & True         & True     & True \\
                & $8^{\prime}$ & No detection & True     & True \\
\cutinhead{Scattered (Gaussian) sources}
                & $2^{\prime}$ & Marginal     & True     & True \\
Off-axis angle (arcmin):  & $5^{\prime}$ & Marginal     & True     & True \\
                & $8^{\prime}$ & No detection & True     & True \\
\cutinhead{Point sources with dust-scattering halo}
                & $2^{\prime}$ & True         & True     & True \\
Off-axis angle (arcmin):  & $5^{\prime}$ & Marginal     & True     & True \\
                & $8^{\prime}$ & No detection & Marginal & True \\
\enddata
\end{deluxetable}
 
As Table~\ref{tab:dust} shows, both the Gaussian and halo dust-scattered models had low-flux ($3.3\times10^{-14}$ \fluxcgs), high-off-axis-angle ($\theta\geq5^{\prime}$) sources for which the \emph{True} point-source classification was reduced to \emph{Marginal}. Higher-flux sources were still classified as True (except for the halo source at $\theta=8^{\prime}$), so even the very strong dust scattering assumed by these models does not significantly degrade the detection significance of many sources.

\section{Blind Survey: Methods and Additional Results}\label{sec:AD}

As described in \S\ref{subsec:survey}, the CLS has no independent set of ground-truth sources with which to directly compare the completeness and reliability of \wavlex and the CSC. We therefore conducted a blind survey in which ten astronomers visually rated candidate sources from 1 (very unlikely to be real) to 5 (very likely to be real) as an independent benchmark. Because visual ratings are subjective, we use them only to compare the two catalogs relative to one another, not to set absolute reliabilities. This appendix describes how the survey was constructed, how the respondents' ratings were weighted and validated, and the survey results not discussed in \S\ref{subsec:survey}.

\begin{figure}
    \centering
    \includegraphics[width=1\linewidth]{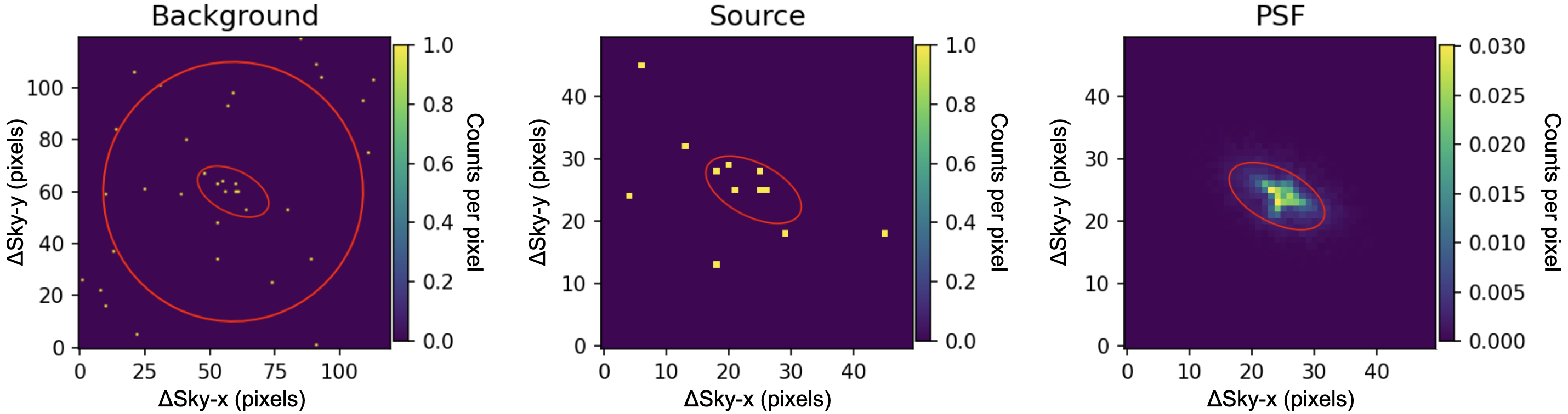}
    \caption{Example source candidate as shown to respondents in the survey described in \S\ref{subsec:survey}. Three "postage-stamp" images show the source (center), the surrounding background (left), and the simulated PSF (right). The source and background regions from \wavlex are outlined in red, and the color scale shows the number of counts per pixel. This source candidate has $\mathcal{L}=6.96$ (\wavlex) and $\mathcal{L}=17.57$ (CSC), with a mean survey rating of $2.5\pm0.9$ (indicating a "Marginal" classification).} \label{fig:stamp}
\end{figure}

To test the detection methods as well as the accuracy and consistency of the respondents' ratings, we selected 100 candidate sources with a wide variety of characteristics. These randomly selected sources comprised: 45  True or Marginal sources from the broad-band catalogs of the CSC and \wavlex, spanning a wide range of likelihood, flux, and $\theta$; 10 duplicates from the latter group, reoriented and randomly distributed to test the consistency of the scores; 5 soft-band and 5 hard-band counterparts of broad-band sources; 10 sources detected only by the CSC and not by \wavlex; 10 sources detected only by \wavlex and not by CSC; and 15 sources simulated with MARX as a "ground-truth" verification set. Each source candidate was presented in a set of three images: the source along with the surrounding background (for context), a zoom-in image of the source, and the simulated PSF at the source location. An example candidate as it appeared in the survey is shown in Figure~\ref{fig:stamp}. The scores were then compared to the relative likelihoods of the sources yielded by \wavlex and CSC.

\subsection{Weighting the respondents' ratings}\label{subsec:AD_A}

Respondents differed in how leniently they used the 1--5 scale, how strongly their ratings distinguished between source classes, and how consistent their ratings were, so an unweighted mean gives a biased or inconsistent respondent the same influence as a more reliable one and inflates the scatter in the ratings for each source. We therefore weight respondents by their reliability, based on their consistency in rating similar sources and on how much their ratings deviated from others.

To determine the weights, each rating from each respondent is modeled with Equation~\ref{rater-model}:
\begin{equation}
  r_{ij} = a_i + \lambda_i\Theta_j +\mathcal{N}(0,\psi_i),
  \label{rater-model}
\end{equation}
Where $r_{ij}$ is the rating of source $j$ by respondent $i$, and each source is assumed to have a latent consensus rating $\Theta_j$. Each respondent's ratings are assumed to follow $\Theta_j$ linearly, with $a_i$ an offset (how lenient they are overall) and $\lambda_i$ a scaling (how strongly their ratings change with $\Theta_j$). There is also the inconsistency of each respondent, modeled as a Gaussian $\mathcal{N}(0,\psi_i)$ with variance $\psi_i$. The offsets $a_i$, scalings $\lambda_i$, and noise variances $\psi_i$ are fit jointly to all respondents' ratings using the expectation-maximization (EM) algorithm \citep{rubin1982em}. Each respondent is then weighted by $\lambda_i^2/\psi_i$, representing the signal-to-noise of their ratings \citep{ferrer2019}, normalized over all respondents using Equation~\ref{rater-weights}:
\begin{equation}
  w_i = \frac{\lambda_i^{2}/\psi_i}{\sum_{k=1}^{10}\lambda_k^{2}/\psi_k}.
  \label{rater-weights}
\end{equation}

Note that $a_i$ is not used in the weighting, since this survey is intended to determine the relative likelihood of sources. The weighted mean rating of each source follows Equation~\ref{weighted-mean}:
\begin{equation}
  \hat{\mu}_j = \frac{\sum_{i=1}^{10} w_i\, r_{ij}}{\sum_{i=1}^{10} w_i}.
  \label{weighted-mean}
\end{equation}

\begin{figure}
    \centering
    \includegraphics[width=1\linewidth]{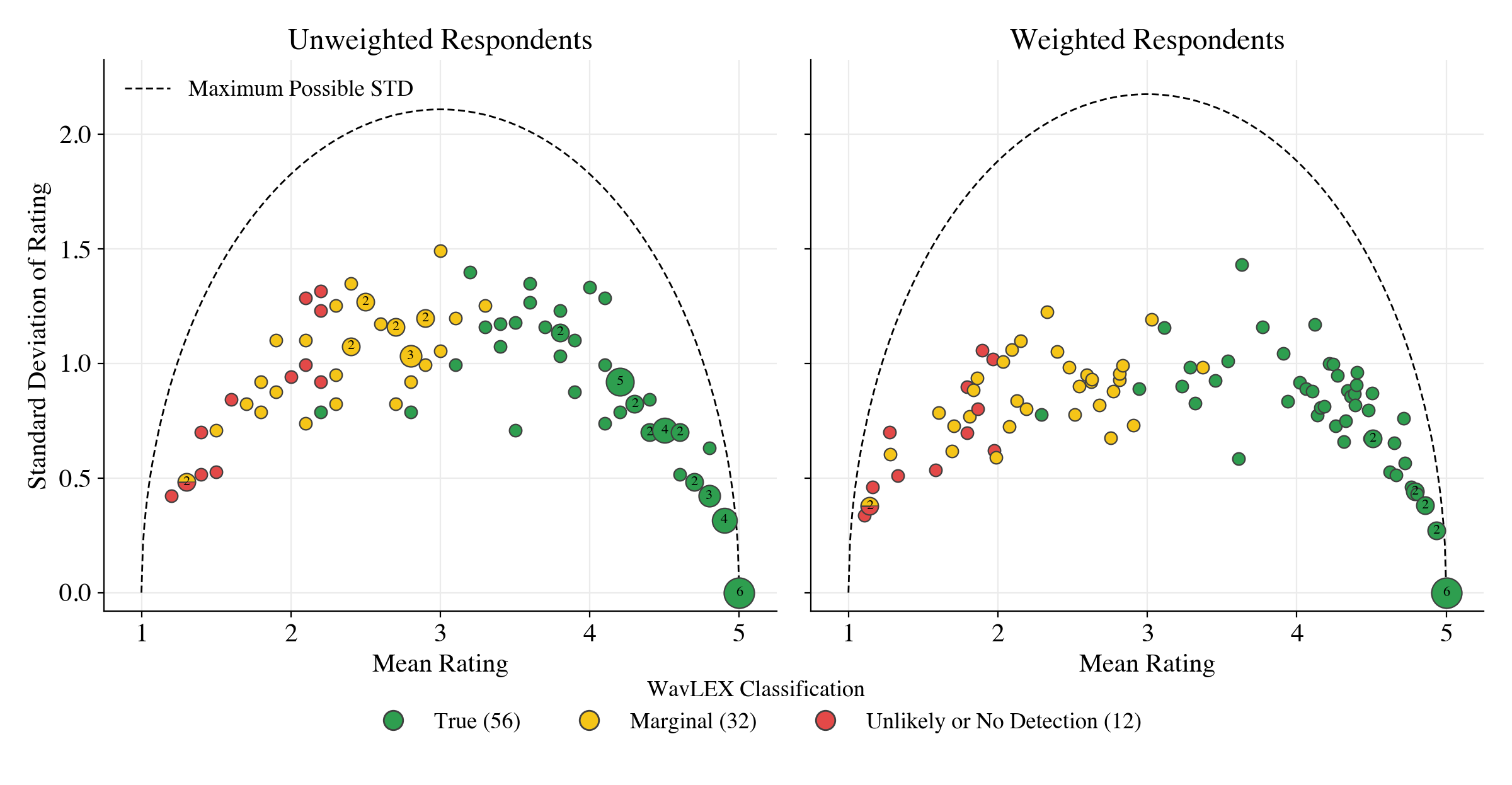}
    \caption{Standard deviation versus mean rating of 100 sources using unweighted (left) and weighted (right) respondent ratings. The dashed line shows the maximum possible standard deviation for the corresponding mean rating. Colors represent \wavlex classification. Larger circles represent multiple sources; the number of sources represented is printed inside each marker.}
    \label{fig:weighted_scores}
\end{figure}

Respondents whose ratings closely track the underlying $\Theta_j$ and are the least affected by noise $\psi_i$ are weighted highest. The weights reward consistent respondents who typically agree with other respondents, but a consistent respondent whose ratings do not scale with $\Theta_j$ is down-weighted (e.g., someone who rates all sources as very likely real). Weighting reduces the standard deviation of the ratings for each source and separates the source classes more clearly, as shown in Figure~\ref{fig:weighted_scores}.

The mean standard deviation decreased from 0.86 to 0.75 after incorporating weights, primarily for sources with high disagreement in their ratings, such as those near a mean rating of 3. The weighted mean ratings also appear better correlated with the likelihood of sources, with weighted mean ratings $< 2$ for all unlikely or undetected sources. Weighting also does not simply let one respondent dominate the ratings: the most reliable respondent was given $25\%$ of the total weight. We therefore adopted the weighted mean as a more rigorous measure of relative source likelihood.

\subsection{Evaluating consistency in respondents' ratings}\label{subsec:AD_B}

\S\ref{subsec:survey} discusses the most significant results of the survey: the correlation between mean source rating and likelihood, for \wavlex vs CSC.  Here we focus on the results from other tests that were incorporated into the visual survey. 

\begin{figure}
    \centering
    \includegraphics[width=1\linewidth]{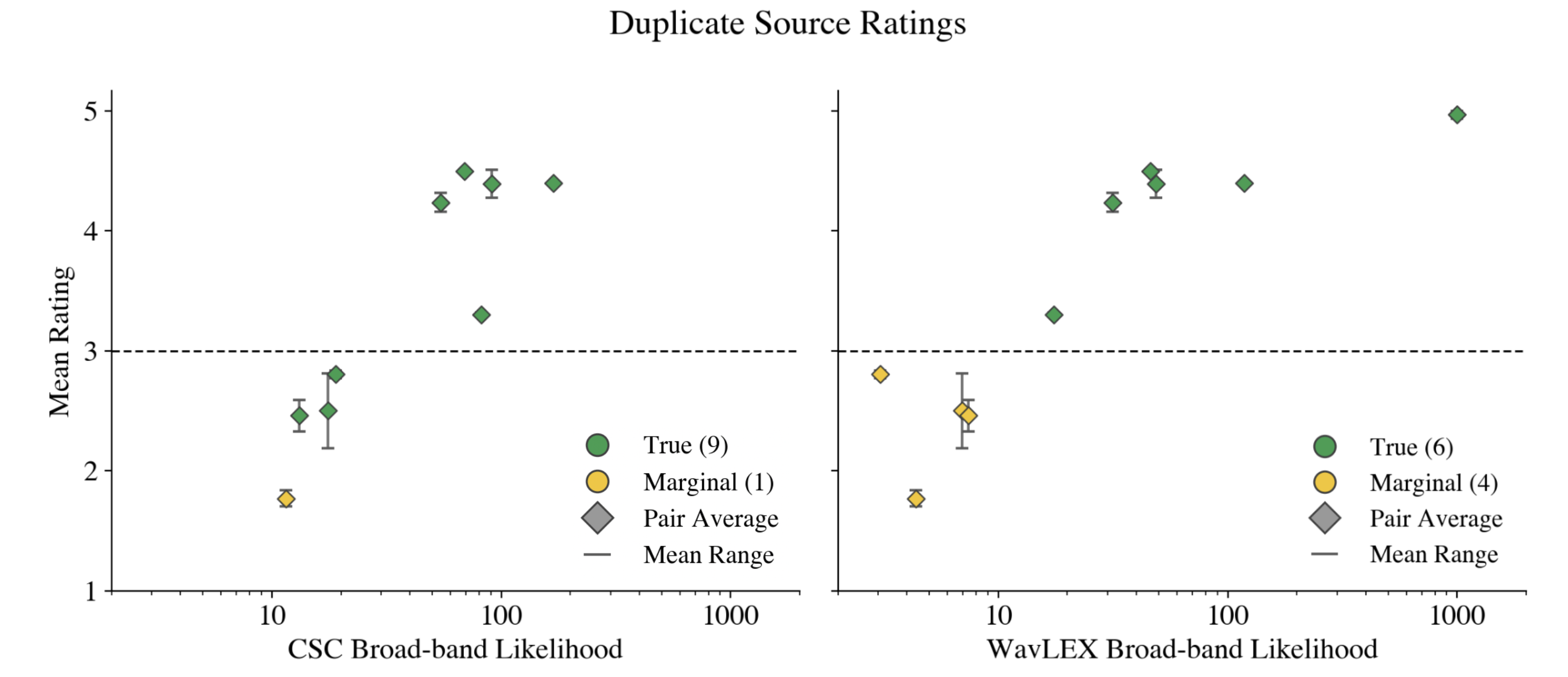}
    \caption{Mean ratings of duplicated sources, spanning a range of likelihood, flux, and off-axis angle $\theta$, versus CSC likelihood (left) and \wavlex likelihood (right). Points are colored by their classification as True or Marginal by the CSC (left) and \wavlex (right). The diamonds show the average combined rating for each pair, and the error bars mark the difference between the two mean ratings. The dashed horizontal line marks a mean rating of 3, where respondents considered the source equally likely to be real or spurious. (The CSC likelihood of one source is out of bounds and is not shown.)}
    \label{fig:duplicates}
\end{figure}

First, as a measure of reliability, we evaluate the consistency of the respondents' ratings. For this purpose, ten sources were each included in the survey twice, appearing in random order and with different orientations to make the duplication less conspicuous. Figure~\ref{fig:duplicates} shows the mean ratings, CSC likelihoods, and \wavlex likelihoods of these sources. The absolute difference between the mean ratings of the duplicate pairs was only 0.19 on average, with a standard deviation of 0.17. This is significantly lower than the average standard deviation between the mean ratings for different source classes, indicating that the respondents' answers were generally self-consistent. We can also use the duplicate pairs to check whether the respondent weights were reasonable. For each respondent we compute the residual sum of squares (RSS) of the differences between their two ratings of each duplicated source. The highest-weighted respondent ($w_1=2$) tied for the lowest RSS ($\mathrm{RSS}=2$), while the lowest-weighted respondent ($w_{10}=0.01$) had $\mathrm{RSS}=34$, close to the $40\pm15$ expected from a respondent rating at random. The weights therefore track self-consistency as intended.

As a final test of reliability, the respondents rated 15 simulated MARX sources with a range of flux and $\theta$. Table \ref{tab:marx_stamps} shows the mean ratings of each individual combination of flux and $\theta$ values. The only sources with mean rating $< 3$ had $Flux=3.3\times10^{-14}$~erg~cm$^{-2}$~s$^{-1}$ and $\theta \geq 8'$; as Figure \ref{fig:limit_heatmaps} shows, these are also the only sources that would fall below the $\mathcal{L}=10$ threshold. Mean ratings correlate with flux and off-axis angle $\theta$, as expected.  

\setlength{\tabcolsep}{10pt}
\begin{deluxetable}{lcccccc}
\tablecaption{Survey results for the 15 MARX simulations as a function of flux and off-axis angle $\theta$. Each cell gives the mean rating of an individual source; the mean ratings as functions of flux and of $\theta$ are shown in the right column and bottom row, respectively. \label{tab:marx_stamps}}
\tablehead{
\colhead{Flux [\fluxcgs]} & \colhead{$\theta = 0.5'$} & \colhead{$\theta = 2'$} & \colhead{$\theta = 5'$} & \colhead{$\theta = 8'$} & \colhead{$\theta = 10'$} & \colhead{$\theta$-averaged mean}
}
\startdata
$3.3\times10^{-13}$ 
& $4.71 \pm 0.76$ & $4.86\pm 0.38$ & $5.00 \pm 0.00$ & $5.00 \pm 0.00$ & $4.48 \pm 0.80$ & $4.81\pm0.54$ \\
$1.0\times10^{-13}$  & $4.72 \pm 0.56$ & $4.76 \pm 0.46$ & $5.00 \pm 0.00$ & $3.77 \pm 1.16$ & $3.94 \pm 0.83$ & $4.44\pm0.84$ \\
$3.3\times10^{-14}$ & $4.24 \pm 1.00$ & $4.62 \pm 0.53$ & $4.26 \pm 0.73$ & $1.79 \pm 0.70$ & $1.87 \pm 0.80$ & $3.36\pm1.46$ \\
\hline
Flux-averaged mean           & $4.56\pm0.80$ & $4.75\pm0.47$ & $4.75\pm0.54$ & $3.52\pm1.54$ & $3.43\pm1.39$ & $4.20\pm1.19$ \\
\enddata
\end{deluxetable}

\subsection{Energy bands vs source ratings}

\begin{figure}
    \centering
    \includegraphics[width=1\linewidth]{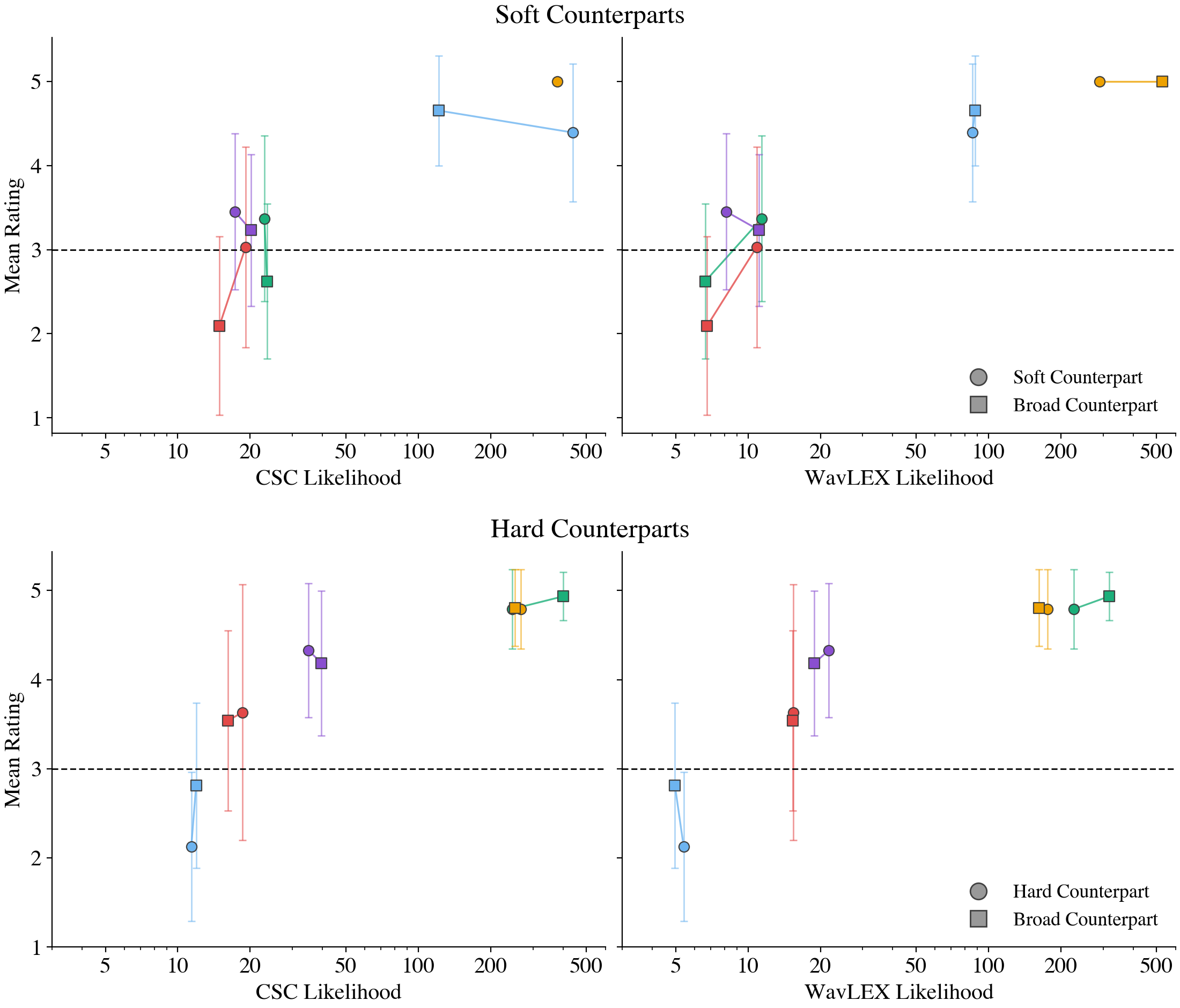}
    \caption{Mean rating versus likelihood for five sources in their broad-band (squares) and soft-band (circles) detections (top), and for five sources in their broad-band (squares) and hard-band (circles) detections (bottom). The horizontal axes show CSC likelihood (left) and \wavlex likelihood (right). Lines connect the two counterparts of each source, and the color of each point identifies the pair to which it belongs. Error bars show the standard deviation of the ratings. The dashed horizontal line marks a mean rating of 3, where respondents considered the source equally likely to be real or spurious.}
    \label{fig:stamp_counterparts}
\end{figure}

Because the CLS catalogs also include soft- and hard-band detections, it is worth examining how the same sources are evaluated in the other energy bands. Figure~\ref{fig:stamp_counterparts} shows the differences in mean rating, CSC likelihood, and \wavlex likelihood for five pairs of soft- and broad-band counterparts (top) and five pairs of hard- and broad-band counterparts (bottom). From the broad-band to the soft-band, the mean rating shifted by $9.2\%\pm16.9\%$, while the CSC likelihood shifted by $48.6\%\pm122.0\%$ and the \wavlex likelihood by $11.2\%\pm52.2\%$. Likelihood and mean rating all increase on average; however, the uncertainties are very large, driven primarily by a few highly variable sources. From the broad-band to the hard-band, the mean rating shifted by $-1.2\%\pm10.4\%$, while the CSC likelihood shifted by $-6.8\%\pm20.4\%$ and the \wavlex likelihood by $0.7\%\pm17.4\%$. The proportional changes in the ratings from the broad-band to the hard-band are, as expected, smaller than for the soft-band; however, the variance in the CSC and \wavlex likelihood changes is consistently greater than that in the mean-rating changes. Direct comparisons using percentage changes are limited because the mean rating is strictly bounded, whereas the likelihood is not. It is also worth noting that likelihood-based metrics evaluated across different energy bands depend heavily on the hardness of each source, not only on systematic differences in detectability (e.g., due to energy-dependent effective area) between the bands.

\section{Medium Band Catalog}\label{sec:AC}
\begin{figure}
    \centering
    \includegraphics[width=1\linewidth]{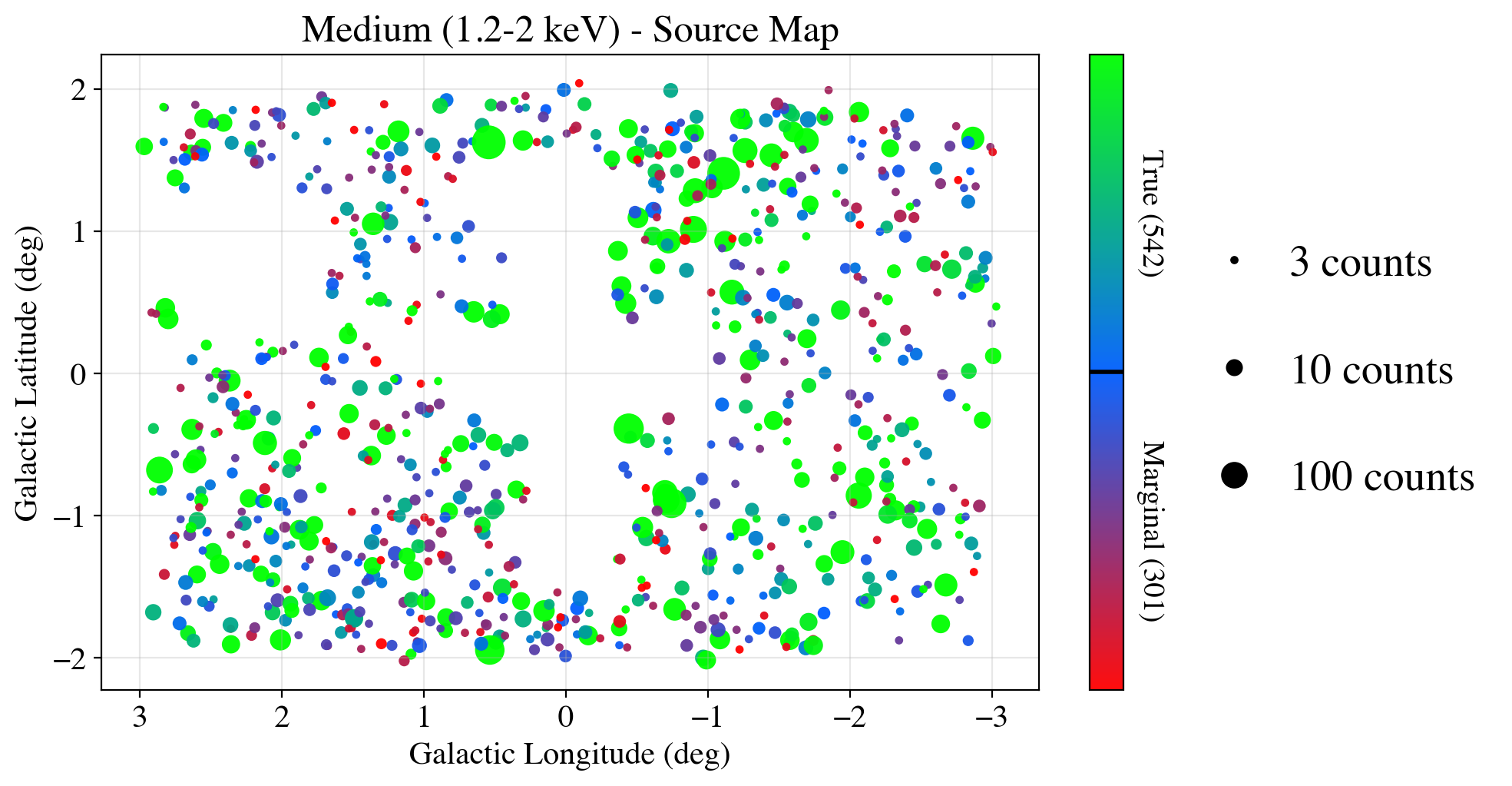}
    \caption{Source map of the CLS catalog in the medium energy band (1.2--2 keV). The color of each source indicates its likelihood: $\mathcal{L}=2.5$ is red, $\mathcal{L}=10$ is blue, and the maximum $\mathcal{L}$ is green. The size of each source scales with the number of counts, as indicated by the legend on the right. The number of True and Marginal sources in each category is also indicated on the color bar.}
    \label{fig:medium_map}
\end{figure}

\begin{figure}
    \centering
    \includegraphics[width=1\linewidth]{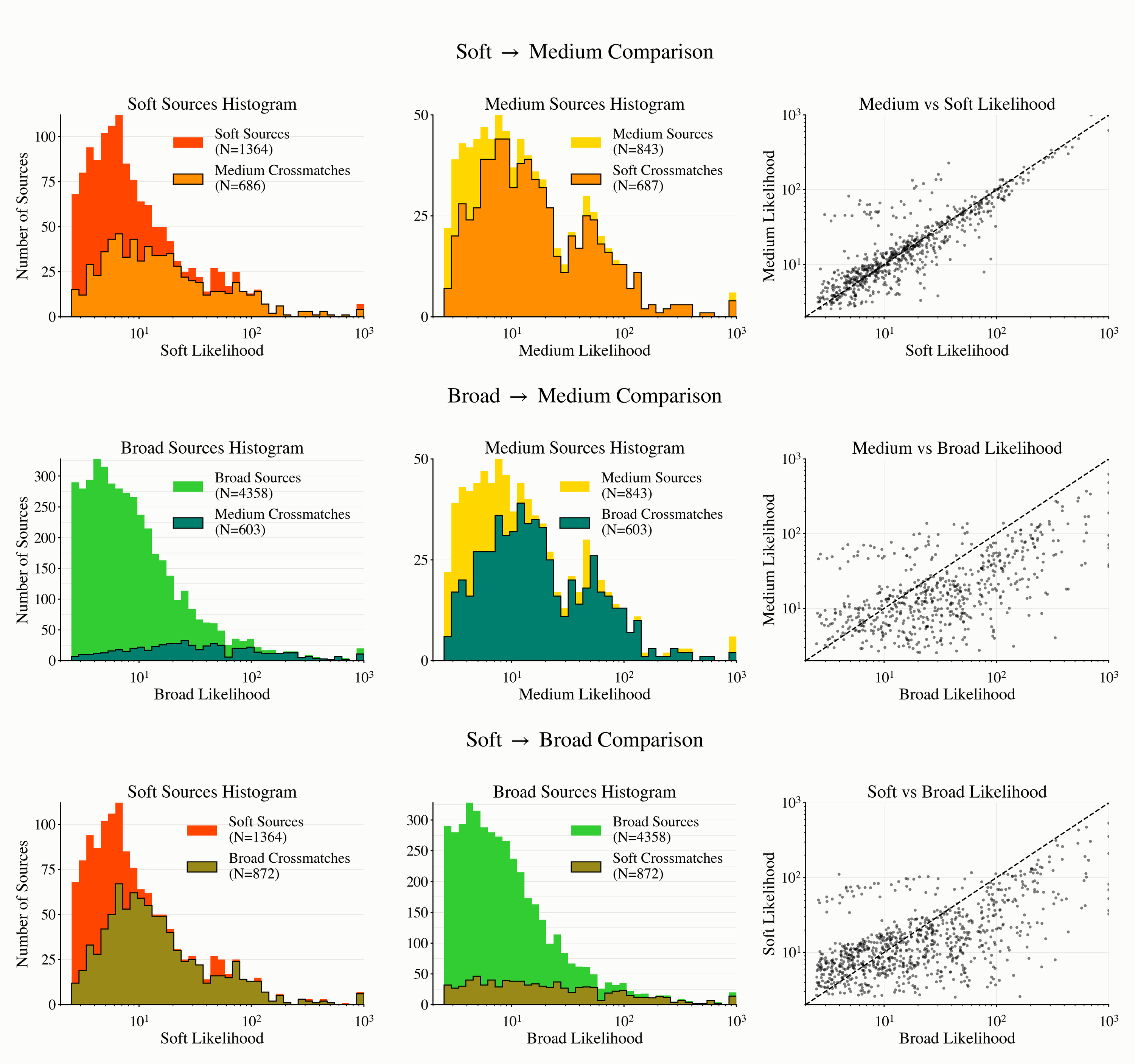}
    \caption{The left and middle columns show the distributions of likelihoods $\mathcal{L}$ for sources in each energy band, and the right column shows scatter plots comparing the likelihoods of crossmatched sources in two energy bands. Each row pairs two energy bands: soft and medium (top), broad and medium (middle), and soft and broad (bottom). In each histogram, the full catalog of one band is overlaid with the subset of its sources crossmatched to the other band using the band-specific likelihood for the x-axis. The energy bands are colored yellow (for medium), red (for soft), and green (for broad). The legends give the group shown in each histogram and the number of sources in that group. There are 40 logarithmically spaced bins for $2.5\leq\mathcal{L}\leq1000$.}
    \label{fig:medium_hist}
\end{figure}

The CLS catalog presented in this paper is divided into three energy bands: soft (0.5--2~keV), hard (2--8~keV), and broad (0.5--8~keV). The boundaries of these energy bands are determined by the limited photon penetration below 2~keV in the GC and the limits of the ACIS-I instrument on \Chandra\ (Figure~\ref{fig:effarea}; \citealt{2017MNRAS.468.2532J}). The CSC broad-band spans 0.5--7~keV, which is similar to the CLS broad-band (0.5--8~keV); however, the CSC also divides the 0.5--2~keV range into two bands: a soft-band (0.5--1.2~keV) and a medium band (1.2--2.0~keV). The boundary at 1.2~keV is motivated by a significant decrease in the ACIS-I effective area below that energy, which makes the 0.5--1.2~keV band proportionally more noise-dominated. For a more direct comparison with the CSC catalog and to better understand how source detection differs in the regime below 2~keV, we used \wavlex\ to create a catalog of sources in the 1.2--2.0~keV medium band. Figure~\ref{fig:medium_map} shows the source map of the medium-band catalog, and Figure~\ref{fig:medium_hist} shows histograms and scatter plots comparing the medium-band catalog to the soft- and broad-band catalogs.


As shown in Figure~\ref{fig:medium_hist}, there are 843 medium-band sources, compared to 1364 soft-band sources. The completion of the medium band within the soft band, 687 sources or $81.5\%$, is the highest of any crossmatched subset, which is expected because the medium band is entirely covered by the soft band. The excess of 521 sources in the soft band could be interpreted as \wavlex accepting spurious detections in the 0.5--1.2~keV regime, which is dominated by noise. However, the completion of the soft band within the broad-band source list ($872/1364$ sources, or $63.9\%$) is only 7.6\% lower than the completion of the medium band within the broad band ($603/843$ sources, or $71.5\%$). This difference could be explained by a population of sources below 1.2~keV that are too soft and faint to be detected in the broad band (where their flux is "diluted" against the background). There is therefore insufficient evidence to suggest that \wavlex is unreliable in the soft band.

In the CLS subset of the CSC, the 1.2--2.0~keV medium band contains 630 sources and the 0.5--1.2~keV soft band contains 132 sources, for a combined 670 individual sources in the 0.5--2.0~keV range. The CSC therefore contains 25.3\% fewer sources than \wavlex in the 1.2--2.0~keV band and 50.9\% fewer in the combined 0.5--2.0~keV range. The larger deficit in the combined range could be explained by the decreased detectability resulting from dividing the energy range into two bands since the CSC does not use a 0.5--2.0~keV band.

\FloatBarrier
\bibliography{cls}
\bibliographystyle{aasjournalv7}

\end{document}